\documentclass[aps, nofootinbib, prd, superscriptaddress, tightenlines,
notitlepage, twocolumn, floatfix]{revtex4-1}
\usepackage{tabularx}
\usepackage{graphicx}
\usepackage{epsfig}
\usepackage{amssymb,bm,tensor}
\usepackage{textcomp}
\usepackage{xcolor}
\usepackage{amsmath}
\usepackage[varg]{txfonts}
\usepackage{enumerate}
\usepackage{mathtools}
\usepackage[normalem]{ulem}
\usepackage{bm} 
\usepackage[OT1]{fontenc}
\usepackage[colorlinks]{hyperref}
\usepackage{multirow}

\def\al{\alpha}
\def\be{\beta}
\def\ga{\gamma}
\def\de{\delta}
\def\ep{\epsilon}

\def\ze{\zeta}
\def\et{\eta}
\def\th{\theta}

\def\io{\iota}
\def\ka{\kappa}
\def\la{\lambda}

\def\rh{\rho}

\def\si{\sigma}

\def\ta{\tau}

\def\ph{\phi}
\def\vp{\varphi}

\def\om{\omega}

\def\Ph{\Phi}
\def\Ps{\Psi}
\def\Om{\Omega}

\def\mn{{\mu\nu}}

\def\lsim{\mathrel{\rlap{\lower4pt\hbox{\hskip1pt$\sim$}}
    \raise1pt\hbox{$<$}}}
\def\gsim{\mathrel{\rlap{\lower4pt\hbox{\hskip1pt$\sim$}}
    \raise1pt\hbox{$>$}}}
\def\sqr#1#2{{\vcenter{\vbox{\hrule height.#2pt
         \hbox{\vrule width.#2pt height#1pt \kern#1pt
         \vrule width.#2pt}
         \hrule height.#2pt}}}}

\def\prt{\partial}
\def\lrpartial{\raise 1pt\hbox{$\stackrel\leftrightarrow\partial$}}

\def\part2{\partial_\alpha \partial^\alpha}

\def\xx'{|\vec x -\vec x'|}

\def\b2{b^\al b_\al}

\newcommand{\beq}{\begin{equation}}
\newcommand{\eeq}{\end{equation}}
\newcommand{\bea}{\begin{eqnarray}}
\newcommand{\eea}{\end{eqnarray}}
\newcommand{\bit}{\begin{itemize}}
\newcommand{\eit}{\end{itemize}}
\newcommand{\rf}[1]{(\ref{#1})}

\newcommand{\Reply}[1]{#1}
\newcommand{\KIAA}{\affiliation{Kavli Institute for Astronomy and
Astrophysics, Peking University, Beijing 100871, China}}
\newcommand{\DoA}{\affiliation{Department of Astronomy, School of Physics,
Peking University, Beijing 100871, China}}

\newcommand{\NAOC}{\affiliation{National Astronomical Observatories,
Chinese Academy of Sciences, Beijing 100012, China}}

\begin{document}

\title{Strong-field effects in massive scalar-tensor gravity for slowly
spinning neutron stars \\ and application to X-ray pulsar pulse profiles}
\date{\today}
\author{Rui Xu}\email[Corresponding author: ]{xuru@pku.edu.cn}\KIAA
\author{Yong Gao}\DoA\KIAA
\author{Lijing Shao}\email[Corresponding author: ]{lshao@pku.edu.cn}\KIAA\NAOC

\begin{abstract} 
Neutron stars (NSs) in scalar-tensor (ST) theories of gravitation can
acquire scalar charges and generate distinct spacetimes from those in
General Relativity (GR) through the celebrated phenomenon of spontaneous
scalarization. Taking on an ST theory with the mass term of the scalar
field, we determine the theory parameter space for spontaneous
scalarization by investigating the linearized scalar field equation. Then
the full numerical solutions for slowly rotating NSs are obtained and
studied in great detail. The resulted spacetime is used to calculate
test-particle geodesics. The lightlike geodesics are used to construct the
profile of X-ray radiation from a pair of hot spots on the surface of
scalarized NSs, which potentially can be compared with the data from the
Neutron star Interior Composition Explorer (NICER) mission for testing the
ST theory.
\end{abstract}

\maketitle

\section{Introduction}
\label{sec:intro}


Gravitational effects are solely described by the metric tensor in General
Relativity (GR). The simplest extension to its field content is adding in a
real scalar field, forming a scalar-tensor (ST) theory of gravitation
\cite{Fujii:2003pa, Will:2018bme, Damour:1992we}. Depending on how the
scalar couples with the metric as well as conventional matters, the
solutions of an ST theory can be identical to or very different from those
of GR. It is particularly interesting to study the ST theories that produce
identical or close enough solutions to those of GR in the weak-field regime
so that they pass all the Solar-system tests as GR does, but that become
sufficiently distinct from GR in the strong-field regime to compete with
it~\cite{Will:2018bme, Berti:2015itd}. In this work we will focus on
neutron stars (NSs) in this kind of ST theories. One well-studied class of
ST theories are characterized by the phenomenon called {\it spontaneous
scalarization}, first discovered and explored by Thibault Damour and Gilles
Esposito-Far\`{e}se for NSs~\cite{Damour:1993hw, Damour:1996ke,
EspositoFarese:2004cc, Damour:2007uf}.

Spontaneous scalarization for NSs can be physically understood as a result
of {\it phase transition}, where for example, the control parameter of the
system can be taken as the baryonic mass of the NS~\cite{Damour:1996ke,
EspositoFarese:2004cc} and Landau's phase transition theory can be
naturally applied~\cite{Sennett:2017lcx}. From the mathematical point of
view, it corresponds to the parameter space (both of the ST theories and of
the NS under study) in which the system acquires two distinct solutions:
one has the GR metric together with a trivial scalar and the other has a
nontrivial scalar with a metric different from GR. The latter
is energetically favored thus represents the physical solution.

The existence of spontaneous scalarization requires conditions on both the
ST theory and the NS. First, the theory must possess nonminimal couplings between the scalar field and the metric. Then, there are restrictions on
the theory parameters and the system parameters. These two points are well
illustrated in Refs.\ \cite{Damour:1993hw, Damour:1996ke} with numerical
solutions to the widely known ST theory proposed by Damour and
Esposito-Far\`{e}se (hereafter, DEF theory). The nonminimal coupling in the
Jordan frame can be transformed to a conformal coupling in the Einstein
frame between the metric and matter fields (see e.g.\
Ref.~\cite{Damour:2007uf}). In the DEF theory, the conformal coupling is
described by an exponential function of the square of the scalar field,
namely $A(\varphi) = \exp(\beta \varphi^2/2)$. The exponential function
decays as the scalar field increases due to a negative coefficient $\be$ in
front of the square of the scalar field. The theory parameter $\be$ needs
to be less than about $-4$ to have spontaneous scalarization for NSs, and
simultaneously, for each valid $\be$ the solutions of spontaneous
scalarization occupy an interval of the system parameter, which, when taken
as the compactness of the system ${\cal C}$, is around $1/|\be|$
\cite{Damour:1993hw, EspositoFarese:2004cc, Ramazanoglu:2016kul}.

When the theory parameter $\be < -4$ naturally takes values of order unity,
NSs coincidentally become the easiest objects to scalarize in the DEF
theory as their compactnesses match right to the required system
compactness ${\cal C} \sim 1/|\be|$. This makes the observations of
pulsars, which are magnetized rotating NSs emitting electromagnetic
radiation, perfectly suitable to test such a theory. Based on the
gravitational radiation formulae derived from the post-Newtonian
approximation \cite{Damour:1992we}, as well as numerical descriptions of
scalarized NSs \cite{Damour:1996ke}, observations of decays in pulsar
orbits due to gravitational-wave (GW) damping have been used to constrain
$\be$ \cite{Freire:2012mg, Antoniadis:2013pzd, Shao:2017gwu,
Anderson:2019eay, Zhao:2019suc}. The most stringent constraint coming from
a combination of multiple binary pulsars indicates $\be > -4.35$ for almost
all supranuclear equations of state (EOSs)~\cite{Wex:2014nva, Shao:2016ezh,
Shao:2017gwu, Shao:2019gjj, Anderson:2019eay, Zhao:2019suc}.

As the theory parameter space shrinks significantly for the DEF theory,
considering a massive scalar instead of a massless one as that in the DEF
theory becomes appealing. Due to the quick Yukawa-type decrease of the
scalar field caused by its mass, the scalar contribution to GW radiation is
automatically suppressed in massive ST theories. Hence, the precise
pulsar-timing observations still do not exclude much of the theory
parameter space of massive ST theories yet. In addition, massless ST
theories of such kind have problems to simultaneously account for the right
behavior in cosmology after the matter-dominated era~\cite{Damour:1993id,
Sampson:2014qqa}, while preserving the strong-field scalarization of NSs.
The whole Universe would have been scalarized when the mass of the scalar
is strictly zero, and a massive scalar is a natural saviour to evade the
scalarization of the whole Universe~\cite{Ramazanoglu:2016kul,
Alby:2017dzl, Anson:2019ebp}. \Reply{Moreover, another motivation from
cosmology for considering massive ST theories is that such models are
natural candidates for dark matter as the massive scalar only interacts
with baryonic matter gravitationally (e.g., see Refs.~\cite{Chen:2015zmx,
Morisaki:2017nit}). }

Numerical solutions of spontaneous scalarization for single spherical NSs
have been constructed in the massive version of the DEF theory with a
constant scalar mass in the Einstein frame \cite{Ramazanoglu:2016kul,
Yazadjiev:2016pcb}. As the Jordan frame, where conventional matters do not
couple with the scalar directly, is usually considered as the physical
frame whose metric can be measured by clocks and meter sticks, we consider
a massive ST theory with a constant scalar mass in this frame. The
conformal transformation connecting the two frames depends on the scalar
field, therefore a constant scalar mass in one frame is no longer constant
in the other. We also point out that there is another difference between
the massive ST theory studied in this work and the DEF theory with a mass
term. The function of the scalar that describes the conformal coupling
takes a rational form instead of the exponential form in the DEF theory. This
is an important check of a different choice because, as numerically shown
in Ref.~\cite{Damour:1993hw}, strong-field effects probe a large segment of
the conformal coupling function, not only in the neighbour around $\varphi
\ll 1$.\footnote{\citet{Damour:1993hw} demonstrated this point using
$A(\varphi) = \exp(-3\varphi^2)$ and $A(\varphi) = \cos(\sqrt{6}\varphi)$
as examples.}

Our work contains the full numerical solutions of single scalarized NSs and
an application of the solutions in constructing the X-ray pulse profiles of
slowly rotating scalarized NSs. In solving NSs numerically, we investigate
the linearized scalar field in detail and obtain the theory parameter space
for spontaneous scalarization to happen. This follows the simple model of
the linearized scalar equation in Ref.~\cite{Damour:1993hw}, with which
Damour and Esposito-Far\`{e}se showed that the scalar field is amplified
when $\be$ is negative to explain the occurrence of spontaneous
scalarization. In applying the numerical solutions to the pulse profile of
X-ray pulsars, we review the prescription in Ref.~\cite{Silva:2018yxz} and
generalize their results to any spherical static spacetime. Explicit
examples are illustrated for the radius, mass, moment of inertia of
scalarized NSs, as well as pulse profiles of X-ray pulsars.

The organization of this paper is as follows. The equations to be solved
are derived from the action of the theory in Sec.~\ref{sec:theory} by
putting forward a metric ansatz for a slowly rotating perfect-fluid NS. Section \ref{sec:lin} demonstrates the
occurrence of spontaneous scalarization and settles the valid parameter
space using the linearized scalar equation, while Sec.~\ref{sec:num}
presents the numerical results for the nonlinear problem. Then, in
Sec.~\ref{sec:app1}, geodesics around scalarized NSs are discussed, and in
Sec.~\ref{sec:app2}, lightlike trajectories are used to calculate the X-ray
flux from a pair of hot spots on the surface of a slowly rotating NS following
Ref.~\cite{Silva:2018yxz}. To conclude the paper, a summary is provided in
Sec.~\ref{sec:disc}. Appendix~\ref{app1} exhibits the series expansion of
the linearized scalar equation at the center of the star.

Throughout this work, we use the geometrized unit system where $G = c = 1$
except when the units are written out explicitly, and the convention of the
metric is $(-,+,+,+)$.

\section{Setup of the problem}
\label{sec:theory}

Writing in the Jordan frame, we study the ST theory given by the
action~\cite{Damour:1996ke, Arapoglu:2019mun}
\begin{align}\label{eq:action}
  S=& \frac{1}{16 \pi } \int  d^{4} x
  \sqrt{-\tilde{g}}\left(\tilde{R}-\tilde{g}^{\mu \nu} \partial_{\mu} \Phi
  \partial_{\nu} \Phi-U(\Ph)+\xi \tilde{R} \Phi^{2}\right)
  \nonumber \\
   &+S_{m}\left[\Ps_{m} ; \tilde{g}_{\mu \nu}\right] ,
\end{align}
where the tildes denote the metric and the metric-related quantities in the
Jordan frame, while $\Ph$ is specifically designated as the scalar field in
this frame. Conventional matters are represented by $\Ps_{m}$ collectively
in the matter action $S_m$, which does not contain the scalar field $\Phi$.
To have a massive scalar field, the scalar potential takes the form
\bea
 U(\Ph) = \left( \frac{2\pi}{\la_\Phi} \right)^2 \Phi^2 \,,
\label{scalarpoten}
\eea
where we have omitted higher-order interactions like the $\Ph^4$ term
(e.g., see Refs.~\cite{Staykov:2018hhc, Arapoglu:2019mun}). The constant
$\la_\Phi$ has dimension of length so that the mass of the scalar can be
defined as
\bea
m_\Ph = \frac{h}{\la_\Ph},
\eea 
where $h$ is the Planck constant that has dimension of length squared in
the geometrized unit system. The nonminimal coupling term in
Eq.~(\ref{eq:action}), $\xi \tilde{R} \Phi^{2}$, is taken from inflationary
models where the inflaton is a single scalar with $\xi$ being the dimensionless coupling
constant (e.g., see Refs.~\cite{Salopek:1988qh, Bezrukov:2007ep,
Hertzberg:2010dc}).

The field equations are obtained by taking variations with respect to
$\tilde g_\mn$ and $\Ph$. They are
\bea
\left( 1+\xi \Phi^2 \right) \tilde R_\mn &=&  8\pi  \left( \tilde T_\mn - \frac{1}{2} \tilde g_\mn \tilde T \right)  + \partial_{\mu} \Phi
  \partial_{\nu} \Phi + \frac{1}{2} \tilde g_\mn  U(\Phi)
\nonumber \\
&& + \xi \left( \tilde D_\mu \tilde D_\nu +  \frac{1}{2} \tilde g_\mn \tilde \Box \right) \Phi^2    ,
\label{jefe}
\eea
and
\bea
\left( \tilde \Box + \xi \tilde R \right) \Phi = \frac{1}{2} \frac{dU}{d\Phi} ,
\label{jscalar}
\eea
where the energy-momentum tensor for conventional matters is
\bea
\tilde T_\mn \equiv - \frac{2}{\sqrt{-\tilde g} } \frac{\de S_m}{\de \tilde g^\mn}  ,
\eea
and the d'Alembert operator is $\tilde \Box = \tilde g^\mn \tilde D_\mu
\tilde D_\nu $ with $\tilde D_\mu$ being the covariant derivative associated with the metric $\tilde g_\mn$.

Attempts to solve Eqs. \rf{jefe} and \rf{jscalar} require metric ansatzes
in the Jordan frame. But we prefer to use a metric ansatz in the Einstein
frame, because equations in the Einstein frame are simpler. With the
conformal transformation
\bea
\tilde g_\mn \equiv A^2(\Ph) g_\mn \equiv \frac{1}{1+\xi \Ph^2} g_\mn ,
\label{aeq}
\eea
and a field redefinition of the scalar satisfying
\bea
\left( \frac{d\vp}{d\Phi} \right)^2 \equiv W(\Ph) \equiv \frac{3}{4} \left( \frac{2\xi\Phi}{ 1+\xi \Phi^2} \right)^2 + \frac{1}{2} \frac{1}{ 1+\xi \Phi^2},
\label{weq}
\eea
the action \rf{eq:action} as well as the field equations \rf{jefe} and
\rf{jscalar} can be transformed into the Einstein frame. They read
\bea
S &=& \frac{1}{16 \pi } \int d^{4} x
  \sqrt{-g } \left( R - 2 g^{\mu \nu} \partial_{\mu} \vp
  \partial_{\nu} \vp - V(\vp) \right)
\nonumber \\
&& +S_{m}\left[\Ps_{m} ; A^2(\varphi) g_{ \mu \nu} \right] ,
\label{eact} 
\eea
\bea
R_\mn = 2 \prt_\mu \vp \prt_\nu \vp + \frac{1}{2} g_\mn V(\vp) + 8\pi \left( T_\mn - \frac{1}{2} g_\mn T \right) ,
\label{eneq}
\eea
and 
\bea
\Box \vp = \frac{1}{4} \frac{dV(\varphi)}{d\vp} - 4\pi \frac{d\ln A(\varphi)}{d\vp} T ,
\label{escalar}
\eea
where the scalar potential and the d'Alembert operator in the Einstein
frame are $V(\vp) = A^4(\varphi) U(\Phi(\varphi))$ and $\Box = g^\mn D_\mu
D_\nu$ respectively. The energy-momentum tensor for conventional matters in
the Einstein frame is
\bea
T_\mn \equiv - \frac{2}{\sqrt{-g} } \frac{\de S_m}{\de g^\mn} = A^2 \tilde T_\mn .
\eea

In the above equations, the metric and the metric-related quantities in the
Einstein frame are written without any decoration, and the scalar field in
the Einstein frame is denoted as $\vp$ specifically. For our purpose, we
will employ the Einstein field equations \rf{eneq} rather than \rf{jefe}
for it is simpler, but the scalar equation \rf{jscalar} rather than
\rf{escalar} to avoid solving the relation between $\Ph$ and $\vp$ from the
differential equation \rf{weq}. Such a treatment is proper as long as we
take care of the transformations throughout the calculation.

Before we proceed with a metric ansatz, we point out that the nonminimal
coupling in Eq.~\rf{eq:action} can be matched to that in the DEF theory
when the scalar field is small. In fact, the coupling function $A(\varphi)$
in the DEF theory is \cite{Damour:1996ke}
\bea
A(\vp ) = \exp \left( \frac{1}{2} \be \vp^2 \right) = 1 + \frac{1}{2} \be \vp^2 + O\left(\vp^4\right) ,
\label{defa}
\eea 
while the coupling function $A(\Phi(\varphi))$ defined in Eq. \rf{aeq}
becomes
\bea
A\left( \Ph (\vp) \right) = 1 - \frac{\xi}{2} \Ph^2 + O\left( \Ph^4 \right) = 1 - \xi \vp^2 + O\left( \vp^4 \right) ,
\eea
when the scalar field is small. Therefore, if \cite{Damour:1996ke}
\bea
\xi = - \frac{1}{2} \be ,
\label{destt}
\eea
the theory studied here is equivalent to the massive DEF theory in the
regime of a weak scalar field.

Now following Ref.~\cite{Damour:1996ke}, we use the metric ansatz
\bea
ds^2 = g_\mn dx^\mu dx^\nu 
&=& -e^{\nu(\rh)} dt^2 + e^{\mu(\rh)} d\rh^2 + \rh^2 d\th^2 
\nonumber \\
&& + \rh^2 \sin^2\th \Big( d\ph + \big(\om(\rh, \th) - \Om\big) dt \Big)^2 , 
\label{metric}
\eea
to simplify the field equations \rf{jscalar} and \rf{eneq}. The
line element is written in the Einstein frame with coordinates $(t, \, \rh,
\, \th, \, \ph)$ and unknown functions $\mu(\rh), \, \nu(\rh), \,
\om(\rh,\th)$. The metric components in the Jordan frame can be obtained
via Eq. \rf{aeq}. The angular velocity of the star, $\Om$, assumed to be
constant, is introduced as the asymptotic value of $\om(\rho, \theta)$ when
$\rho \to \infty$.

The fluid variables, on the other hand, are conventionally written in the
Jordan frame, namely that the energy-momentum tensor in the Jordan frame
takes the form
\bea
\tilde T^\mn = (\tilde \ep + \tilde p) \tilde u^\mu \tilde u^\nu + \tilde g^\mn \tilde p ,
\eea
where $\tilde \ep$ and $\tilde p$ are related by the EOS, and the
4-velocity $\tilde u^\mu$ is
\bea
\tilde u^\mu = \frac{1}{ \sqrt{-\tilde g_{tt} - 2\Om \tilde g_{t\ph} - \Om^2 \tilde g_{\ph\ph} } } \left( 1, 0, 0, \Om \right) .
\eea

Confined to slowly rotating NSs and keeping only the linear terms of $\om$
and $\Om$ in the field equations, we find that the fluid variables, $\tilde
\ep, \, \tilde p$, and the scalar field, $\Ph$, are functions of the radial
coordinate $\rh$ alone~\cite{Hartle:1967he}, and that $\om(\rh,\th)$
remarkably obeys an equation in the same form as that in the massless case
in Ref.~\cite{Damour:1996ke}:
\bea
&& \frac{1}{\rh^4} e^{ -\frac{\mu-\nu}{2} } \frac{\partial }{\partial \rh}\left(\rh^4 e^{ -\frac{\mu+\nu}{2} } \frac{\partial \om}{\partial \rh}\right)+\frac{1}{\rh^2 \sin ^3\th}  \frac{\partial }{\partial \th}\left(\sin ^3\th \frac{\partial \om}{\partial \th}\right) 
\nonumber \\
&& =  \frac{16\pi G}{c^4}  A^4 (\tilde \ep + \tilde p) \om ,
\label{omeq}
\eea
with the coupling function $A$ defined in Eq.~\rf{aeq} and no contribution from the scalar potential $U$. Equation \rf{omeq}
allows a separation of variables and additionally implies that $\om$ is
independent of $\th$ when the asymptotic behavior of $\om$ is taken into
account \cite{Hartle:1967he, Poisson:2014}. Therefore, $\om$ is also a
function of the radial coordinate $\rh$ alone, and the field equations
\rf{jscalar} and \rf{eneq} become a group of ordinary differential
equations (ODEs).

With the change of variable
\bea
e^\mu \equiv \left( 1 - \frac{2m(\rh)}{\rh} \right)^{-1} ,
\eea
the ODEs written out explicitly are
\begin{align}
m' =& 4\pi  \rh^2 A^4 \tilde \ep + \frac{1}{2} \rh (\rh - 2 m ) W \Ph^{\prime\,2} + \frac{1}{4} \rh^2 A^4 U ,
\nonumber \\
\nu' =& \frac{ 8\pi \rh^2 A^4 \tilde p }{\rh - 2m} + \frac{2m}{\rh(\rh-2m) } + \rh W \Ph^{\prime\,2} - \frac{1}{2} \frac{\rh^2}{ (\rh - 2 m)} A^4 U  ,
\nonumber \\
\om'' =&  \frac{4\pi \rh}{\rh - 2m} A^4 (\tilde \ep + \tilde p) ( \rh \om' + 4\om )  +  \left(  \rh W \Ph^{\prime\,2} - \frac{4}{\rh} \right) \om' ,
\nonumber \\
\tilde p' =& - (\tilde \ep + \tilde p) \left(  \frac{4\pi \rh^2 A^4 \tilde p }{\rh - 2m} + \frac{m}{\rh(\rh-2m) } \right.
\nonumber \\
&  \left. + \frac{1}{2} \rh W \Ph^{\prime\,2} + \frac{A'}{A} - \frac{1}{4} \frac{\rh^2}{ (\rh - 2 m)} A^4 U \right) ,
\nonumber \\
\Ph'' =&  \frac{ 4\pi \rh A^4}{\rh - 2m} \left( \rh (\tilde \ep - \tilde p) \Ph'  + \frac{1}{W A} \frac{d A}{d\Ph} (\tilde \ep - 3 \tilde p) \right)  
\nonumber\\
&  - \frac{ W' \Ph^{\prime} }{2 W } - \frac{2(\rh - m) }{\rh (\rh - 2m) } \Ph' 
\nonumber \\
&  + \frac{\rh A^4}{\rh - 2m} \left( \frac{1}{2} \rh U \Ph' + \left( \frac{U}{A}\frac{d A}{d\Ph}  + \frac{1}{4} \frac{dU}{d\Ph} \right) \frac{1}{W} \right) , 
\label{odegroup}
\end{align}
where the primes denote derivatives with respect to the radial coordinate
$\rh$, and $U, \, A$ and $W$ are functions of $\Ph$ given in
Eqs.~\rf{scalarpoten}, \rf{aeq} and \rf{weq}. We point out that the above
set of equations recover those in Refs.~\cite{Damour:1996ke,
Ramazanoglu:2016kul, Yazadjiev:2016pcb, Arapoglu:2019mun, Staykov:2018hhc}
if $U, \, A$ and $W$ take appropriate forms corresponding to the coupling
functions and the scalar potentials there, and the transformation between
the Einstein frame and the Jordan frame is accounted for properly. The
group of equations in \rf{odegroup} is completed by adding in the EOS of
NSs.

A preliminary inspection of the equations in \rf{odegroup} reveals that the
coupled equations are those of $m, \, \tilde p$ and $\Ph$. The function
$\nu$ can be integrated out with the condition of asymptotic flatness once
$m, \, \tilde p$ and $\Ph$ are known. The function $\om$ can be scaled by
an arbitrary factor as its equation is homogeneous. In addition, to
guarantee $\om''|_{\rh = 0}$ to be finite, we have $\om'|_{\rh = 0} = 0$.
Therefore, numerical solutions to the group of equations in \rf{odegroup}
only rely on the initial conditions of $m, \, \tilde p$ and $\Ph$ at $\rh =
0$, which are
\beq
m|_{\rh = 0} = 0, \quad
\tilde p|_{\rh = 0} = \tilde p_c , \quad
\Ph|_{\rh = 0} = \Ph_c , \quad
\Ph'|_{\rh = 0} = 0 .
\label{inicon}
\eeq   
In conclusion, there are two system parameters, the central pressure $\tilde
p_c$ and the central scalar field $\Ph_c$, that can be adjusted in
searching for numerical solutions of spontaneous scalarization, given a
theory with parameters $\xi$ and $m_\Phi$.
 
From the asymptotic behavior of $m$ and $\om$, two important global
quantities, the ADM mass $M$ and the angular momentum $J$, can be
extracted. Specifically speaking, when $\rh \rightarrow \infty$ we have
\bea
g_{\rh\rh} &=& \left( 1 - \frac{2m(\rh)}{\rh} \right)^{-1} \rightarrow 1 + \frac{2M}{\rh} ,
\nonumber \\
g_{t\ph} &=& (\om - \Om) \rh^2 \sin^2\th  \rightarrow - \frac{2J\sin^2\th}{\rh} ,
\eea  
indicating
\beq
M = m|_{\rh \rightarrow \infty}, \quad
J = \frac{1}{6} \rh^4 \om' |_{\rh \rightarrow \infty} .
\eeq
The moment of inertia of the star can then be calculated via
\bea
I = \frac{J}{\Om} = \frac{1}{6} \frac{\rh^4 \om'}{\om} \Big|_{\rh \rightarrow \infty} .
\eea
We point out that as $\Ph \rightarrow 0$ when $\rh \rightarrow \infty$, the
coupling function $A$ becomes unity so that the Einstein frame and the
Jordan frame coincide. Hence, the ADM mass $M$, the angular momentum $J$,
and the moment of inertia $I$ are independent of the frame choice.

\section{Solutions of spontaneous scalarization}

In this section, we investigate the NS solutions from the action
(\ref{eq:action}) in a linearized limit (Sec.~\ref{sec:lin}) and in the
fully nonlinear problem (Sec.~\ref{sec:num}).

\subsection{Linearized scalar field equation}
\label{sec:lin}

As a warm-up, we deal with the linearized version of the scalar field
equation in \rf{odegroup}. This is sufficient to show how spontaneous
scalarization occurs and yields restrictions on the parameter space of the
theory for it to happen. By taking 
\[
A^2 = 1 - \xi \Ph^2 + O(\Ph^4), \quad W = \frac{1}{2} + O(\Ph^2) \,,
\]
and keeping only the linear terms of $\Ph, \, \Ph'$ and $\Ph''$, the scalar
equation in \rf{odegroup} simplifies to
\bea
\Ph'' &=& \frac{ 4\pi \rh^2}{\rh - 2m}  (\tilde \ep - \tilde p) \Ph'  - \frac{2(\rh - m) }{\rh (\rh - 2m) } \Ph' 
\nonumber \\
&&  - \frac{ 8\pi\xi \rh}{\rh - 2m}  (\tilde \ep - 3\tilde p) \Ph + \left( \frac{2\pi}{\la_\Ph} \right)^2 \frac{\rh }{\rh - 2m} \Ph ,
\label{linph}
\eea
where we assume that $m$, $\tilde \ep$ and $\tilde p$ take GR results and are known. Realistic EOSs of NSs can be used to obtain GR
solutions of $m$, $\tilde \ep$ and $\tilde p$~\cite{Lattimer:2000nx}. For
demonstration purpose, we currently employ a toy EOS
\bea
\tilde \ep = {\rm const.} \,,
\label{toyeos}
\eea
which has the advantage to have analytical expressions of $m$ and
$\tilde p$, in order to show how spontaneous scalarization occurs.

Using a characteristic length $l_0$ to define the dimensionless quantities
\beq
x \equiv \frac{\rh}{l_0}, \quad \!
y \equiv \frac{m}{l_0}, \quad \!
u \equiv 4\pi l_0^2 \tilde \ep, \quad \!
v \equiv 4\pi l_0^2 \tilde p , \quad \!
a \equiv \frac{2\pi l_0}{\la_\Ph} ,
\eeq
turns out to be handy. The dimensionless version of Eq. \rf{linph} can be
written as
\bea
&& \left( 1-\frac{2y}{x} \right) \Ph_{xx} + \left(\frac{2}{x} \left( 1 - \frac{y}{x} \right) - (u-v)x \right) \Ph_x
\nonumber \\
&& + \left( 2\xi(u-3v) - a^2 \right) \Ph = 0 ,
\label{linphdim}
\eea
where the subscript $x$ denotes the derivative with respect to $x$. In the
case of a constant $\tilde \ep$, it is convenient to take the
characteristic length as
\bea
l_0 = \sqrt{\frac{3}{8\pi \tilde \ep} } , 
\eea
and then the GR solution for $m$ and $\tilde p$ can be written as
\bea
&& y = \frac{1}{2} x^3,
\nonumber \\
&& v = \frac{3\et\sqrt{1-x^2} - 1}{2(1-\et\sqrt{1-x^2})} ,
\label{mpsol}
\eea
where $\et$ is an integral constant that fixes the parameters of the star.
For example, the dimensionless radius of the star is
\bea
x_s = \sqrt{ 1 - \frac{1}{9\et^2} },
\eea 
and the compactness of the star is
\bea
{\cal C} = \frac{y}{x} \Big|_{x=x_s} = \frac{1}{2} \left( 1 - \frac{1}{9\et^2} \right). 
\eea
Assuming $\et \in ({1}/{3}, 1)$, the compactness ${\cal C}$ increases with
$\et$ from $0$ to ${4}/{9}$.

Though equipped with analytical expressions of $y, \, u$ and $v$, the
analytical solution to Eq. \rf{linphdim} is difficult to find. Therefore we take a semi-analytical approach. Starting at the center
of the star, for given values of ${\cal C}, \, \xi$ and $a$, we find that
the requirement of a finite $\Ph|_{x=0}$ fixes the solution up to a scaling
constant (see Appendix \ref{app1}). The scaling constant preserves the
ratio of $\Ph_x$ to $\Ph$, so the solution is in fact fully determined once
we assign $\Ph|_{x=0} = 1$ by taking advantage of the homogeneity of
Eq.~(\ref{linphdim}). However, the asymptotic solution of Eq. \rf{linphdim}
clearly takes the general form
\bea
\Ph \rightarrow \frac{\Ph_+}{x} e^{ax} + \frac{\Ph_-}{x} e^{-ax} ,
\label{asyph}
\eea  
where the integral constant $\Ph_+$ must vanish and the integral constant
$\Ph_-$ must be nonzero for nontrivial physical solutions. For given values
of ${\cal C}, \, \xi$ and $a$, this asymptotic condition is generally not
satisfied by the already fixed solution obtained through integrating from
the center, causing the only solution to be the trivial one: $\Ph = 0$
everywhere.

To find for what values of ${\cal C}, \, \xi$ and $a$, there are nontrivial
solutions, we split Eq. \rf{linphdim} into the interior part and the
exterior part that match at the surface of the star where $x=x_s$. The
interior equation is solved from the center straightforwardly, while the
exterior equation is solved from the surface of the star with the shooting
method to guarantee $\Ph \rightarrow 0$ at infinity. In applying the
shooting method, we fix $\Ph|_{x=x_s} =1$ and adjust $\Ph_x|_{x=x_s}$ to
achieve a vanishingly small $\Ph$ at large enough $x$. The interior and the
exterior solutions should match at $x = x_s$, by which we mean that the
value of $\Ph_x/\Ph$ at the surface obtained from the interior
solution should equal to that determined by the shooting method in solving
the exterior equation.

\begin{figure}
  \includegraphics[width=\linewidth]{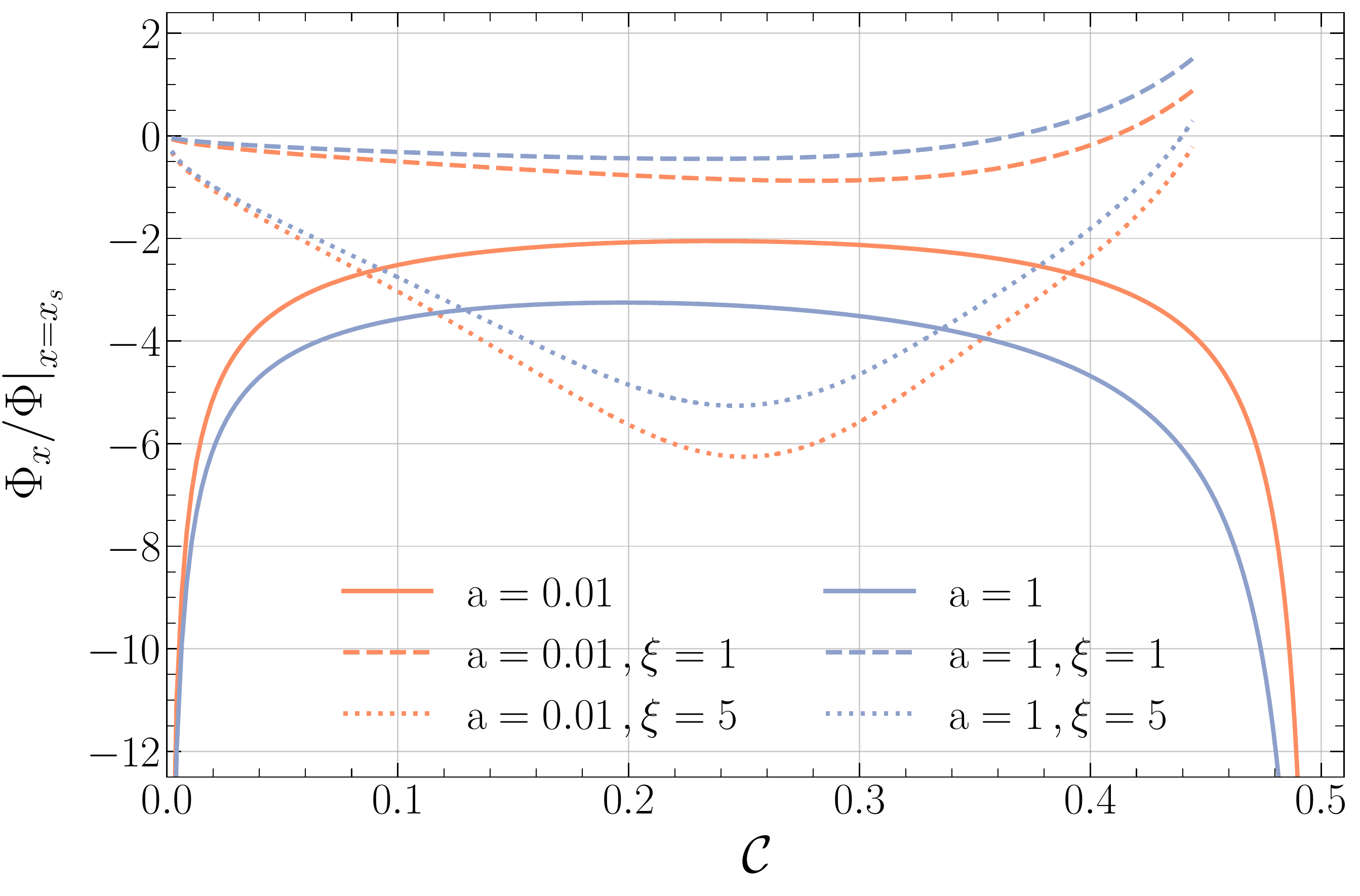}
  \caption{$\Phi_x / \Phi \big|_{x = x_s}$ with respect to the compactness
  ${\cal C}$. The solid curves are for the exterior solutions, while the
  dashed and the dotted curves are for the interior solutions.}
 \label{fig1}
 \end{figure}
 
 \begin{figure}
   \includegraphics[width=\linewidth]{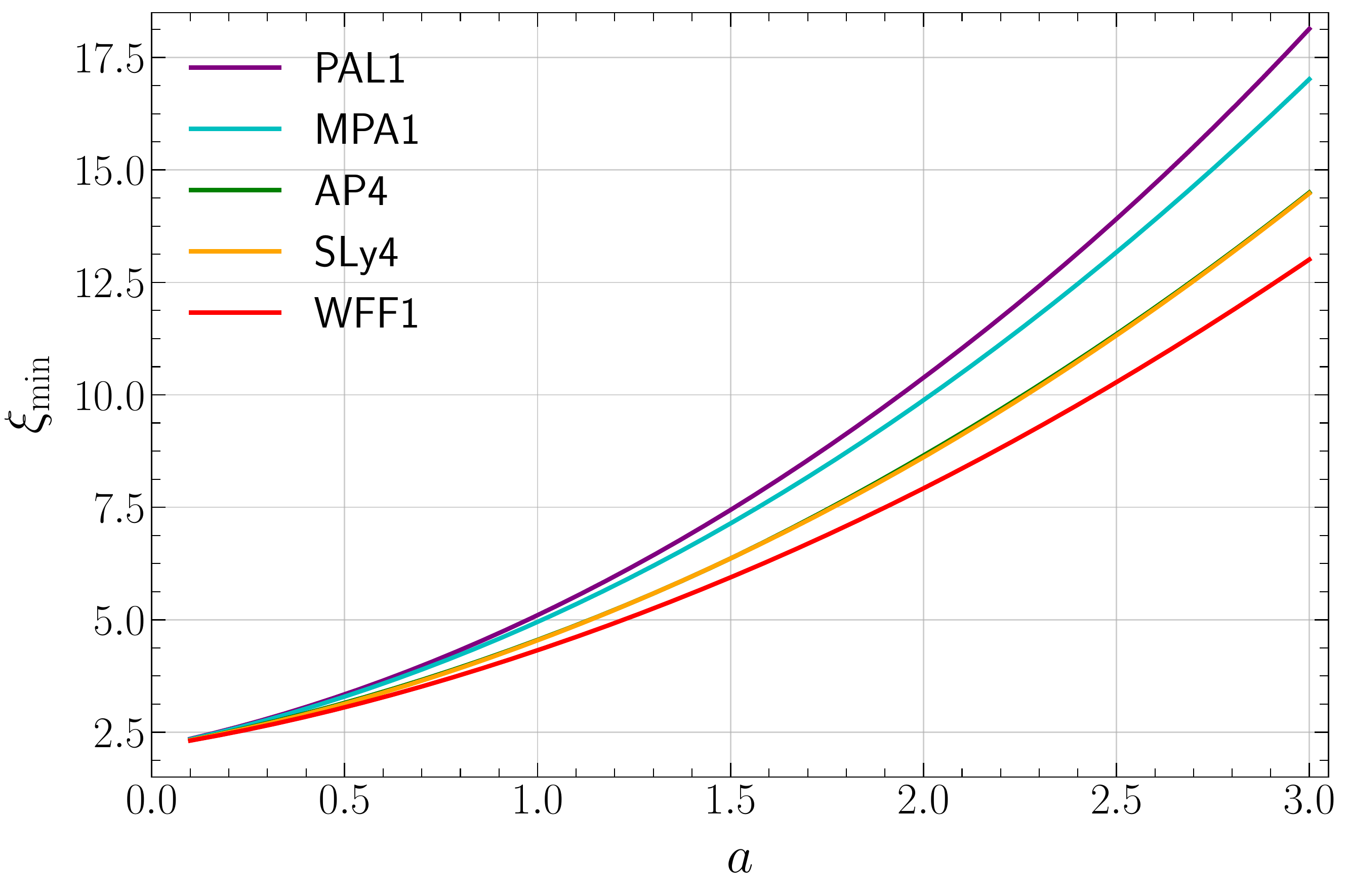}
   \caption{The minimal value of $\xi$ for spontaneous scalarization versus
   the dimensionless scalar mass $a$ for five realistic EOSs; see
   Eq.~(\ref{eq:Phi:mass}) to convert $a$ to the physical scalar mass $m_{\Ph}$. The
   curves for the EOS AP4 and for the EOS SLy4 largely overlap in the
   graph. }
 \label{fig2}
 \end{figure}
 
 \begin{figure*}
  \includegraphics[width=0.7\linewidth]{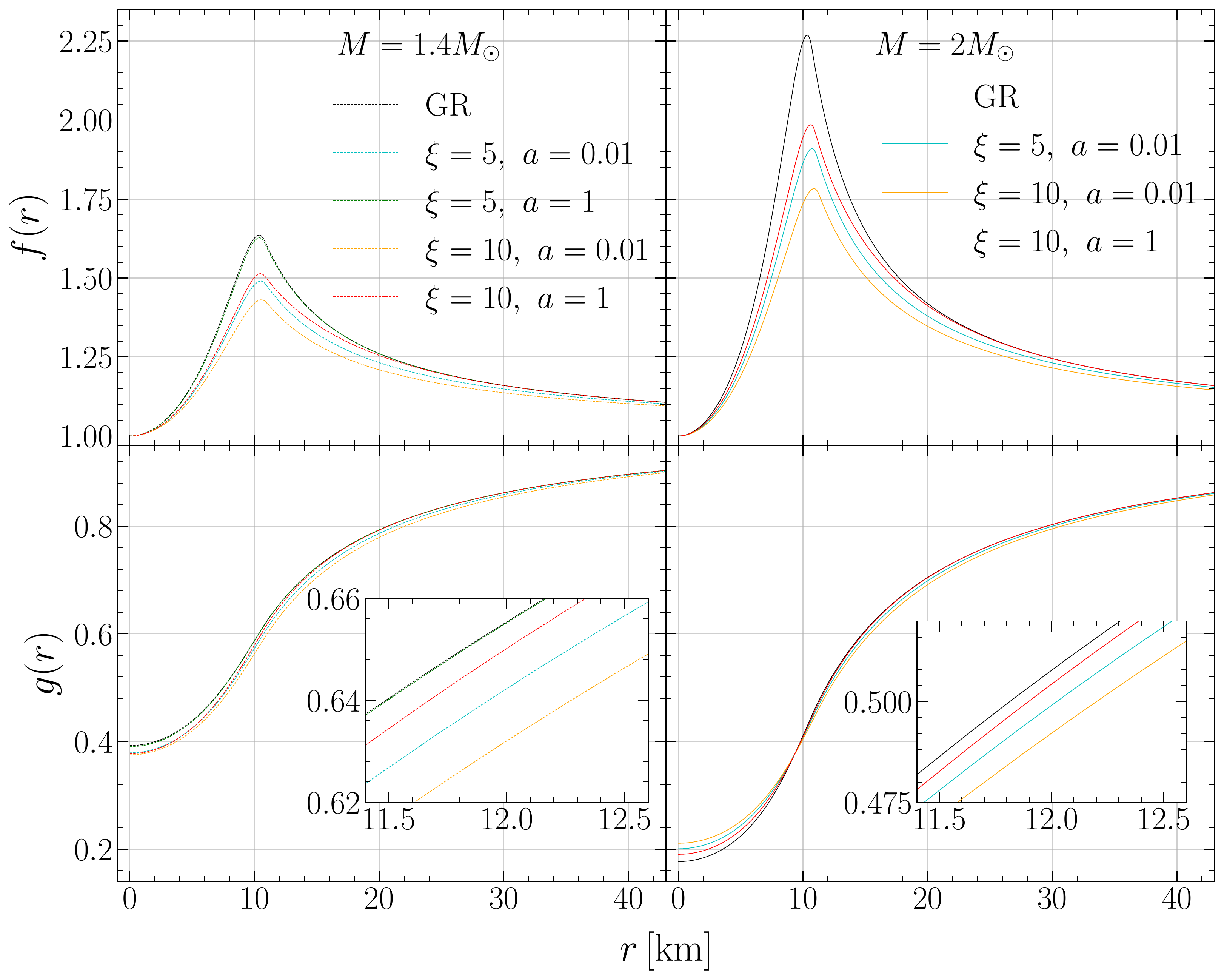}
   \caption{\Reply{ Representative solutions for the Jordan-frame metric
   components $g(r)$ and $f(r)$ defined in Eq.~\rf{tranfg}. EOS AP4 is
   used in calculation, and the ADM masses of the NSs are chosen to be $1.4
   \, M_{\odot}$ (left panels) and $2.0 \, M_{\odot}$ (right panels). } }
 \label{figadd1}
 \end{figure*}
 
\begin{figure*}
  \includegraphics[width=0.7\linewidth]{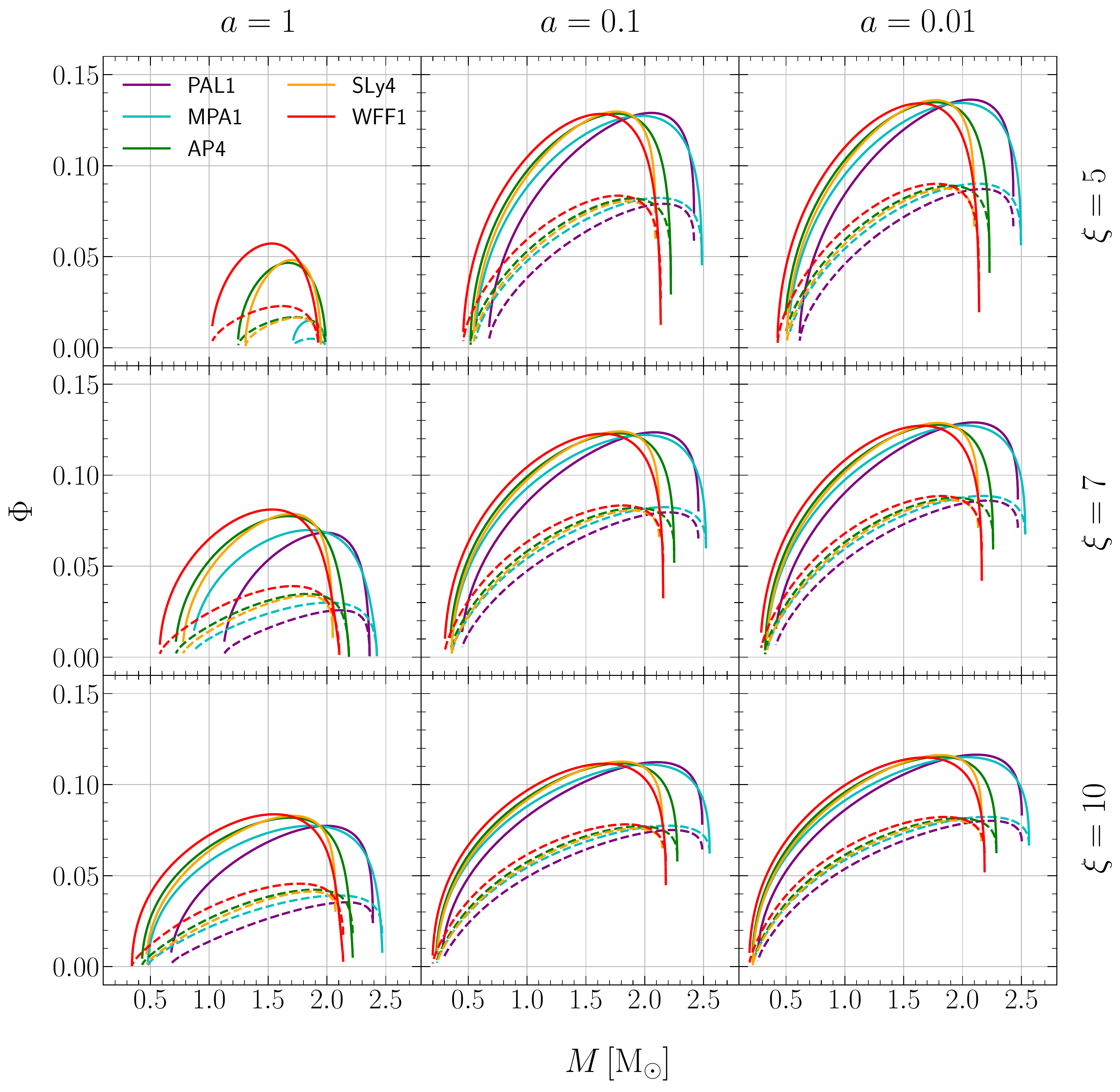}
  \caption{The values of the scalar field at the center (solid curves) and on
  the surface (dashed curves) of the star versus its ADM mass $M$ for three
  typical values of $\xi$ and $a$. The same five EOSs as in Fig. \ref{fig2}
  are used. Note that no solutions of spontaneous scalarization exist with
  the EOS PAL1 for $\xi = 5$ and $a=1$, as $\xi_{\rm min}$ is slightly
  greater than 5 when $a = 1$ for the EOS PAL1 (cf.\ Fig. \ref{fig2}). }
  \label{fig3}
  \end{figure*}

Varying ${\cal C}, \, \xi$ and $a$, values of $\Ph_x/\Ph$ on the surface of
the star for both interior and exterior solutions are calculated
numerically. Figure \ref{fig1} plots these values with reference to the
compactness ${\cal C}$ for two values of $\xi$ and $a$ to show the
following features:
\begin{enumerate}
    \item The existence of nontrivial solutions demands $\xi$ to be greater
    than or equal to a critical value $\xi_{\rm min}(a)$ for a given value
    of $a$.
    \item When $\xi > \xi_{\rm min}(a)$ for a given value of $a$, the
    nontrivial solutions come up at two values of ${\cal C}$.
\end{enumerate} 

The relation $\xi_{\rm min}(a)$ and the two values of ${\cal C}$ for given
$\xi > \xi_{\rm min}(a)$ define the boundaries of the parameter space for
spontaneous scalarization to happen. This is verified in the numerical
results for the nonlinear problem in Sec.~\ref{sec:num}. Figure \ref{fig1}
also suggests that Schwarzschild black holes cannot be scalarized in the
theory studied here since $\Ph_x/\Ph$ at the surface necessary for a
nontrivial physical solution diverges when ${\cal C} \rightarrow {1}/{2}$;
though the EOS in Eq.~\rf{toyeos} does not produce black holes, the exterior
equation of Eq.~\rf{linphdim} still exists if the metric is taken to be the
Schwarzschild metric.

 \begin{figure*}
  \includegraphics[width=0.7\linewidth]{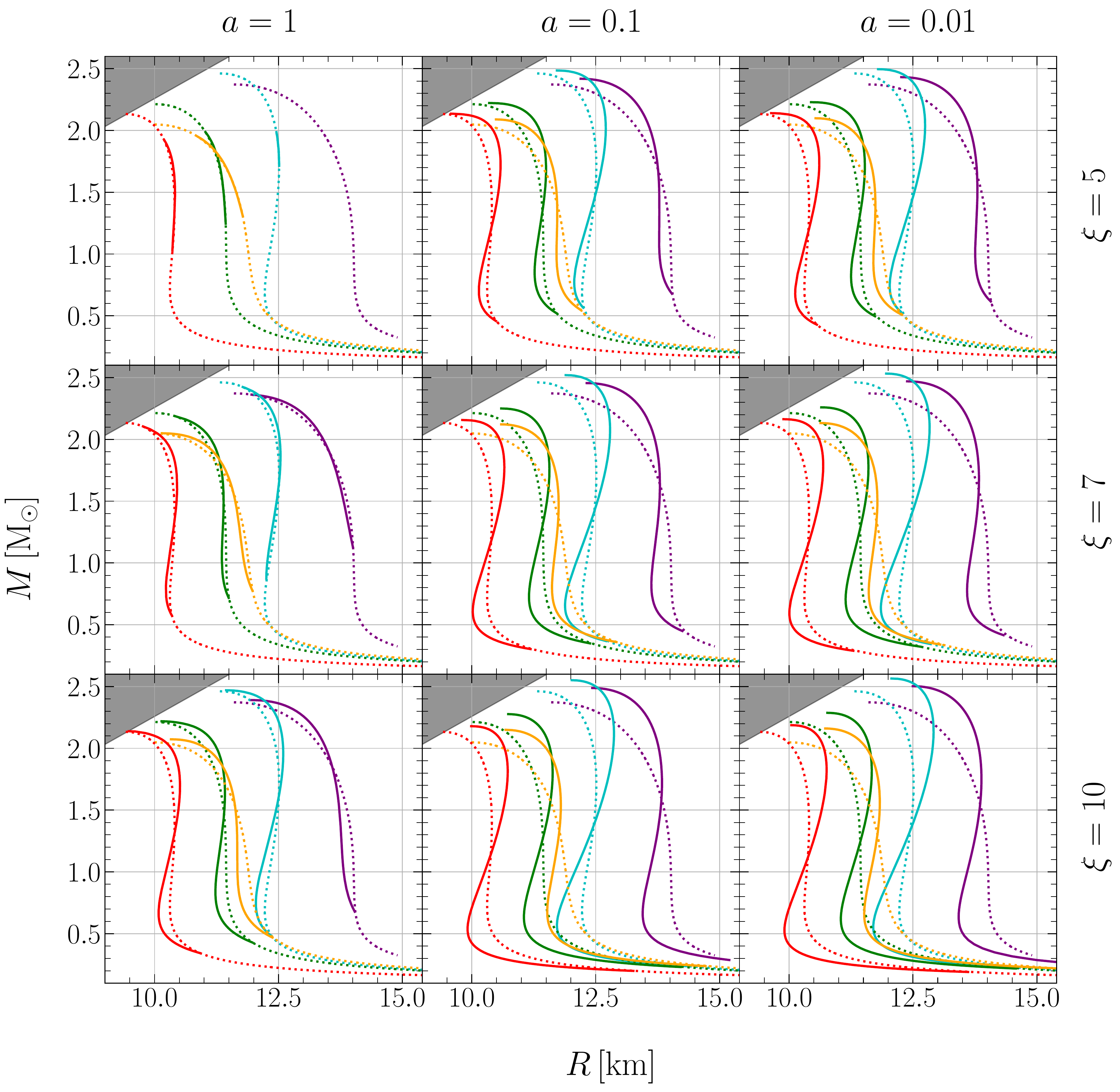}
   \caption{Mass-radius plots for the same values of $\xi$ and $a$, as well
   as the same set of EOSs used in Fig.~\ref{fig3}. The solid curves are
   results of spontaneous scalarization, while the dotted curves are
   results from GR. The shaded region is $ R < 3M$ for guiding purpose. }
 \label{fig4}
 \end{figure*}
 
 \begin{figure*}
  \includegraphics[width=0.7\linewidth]{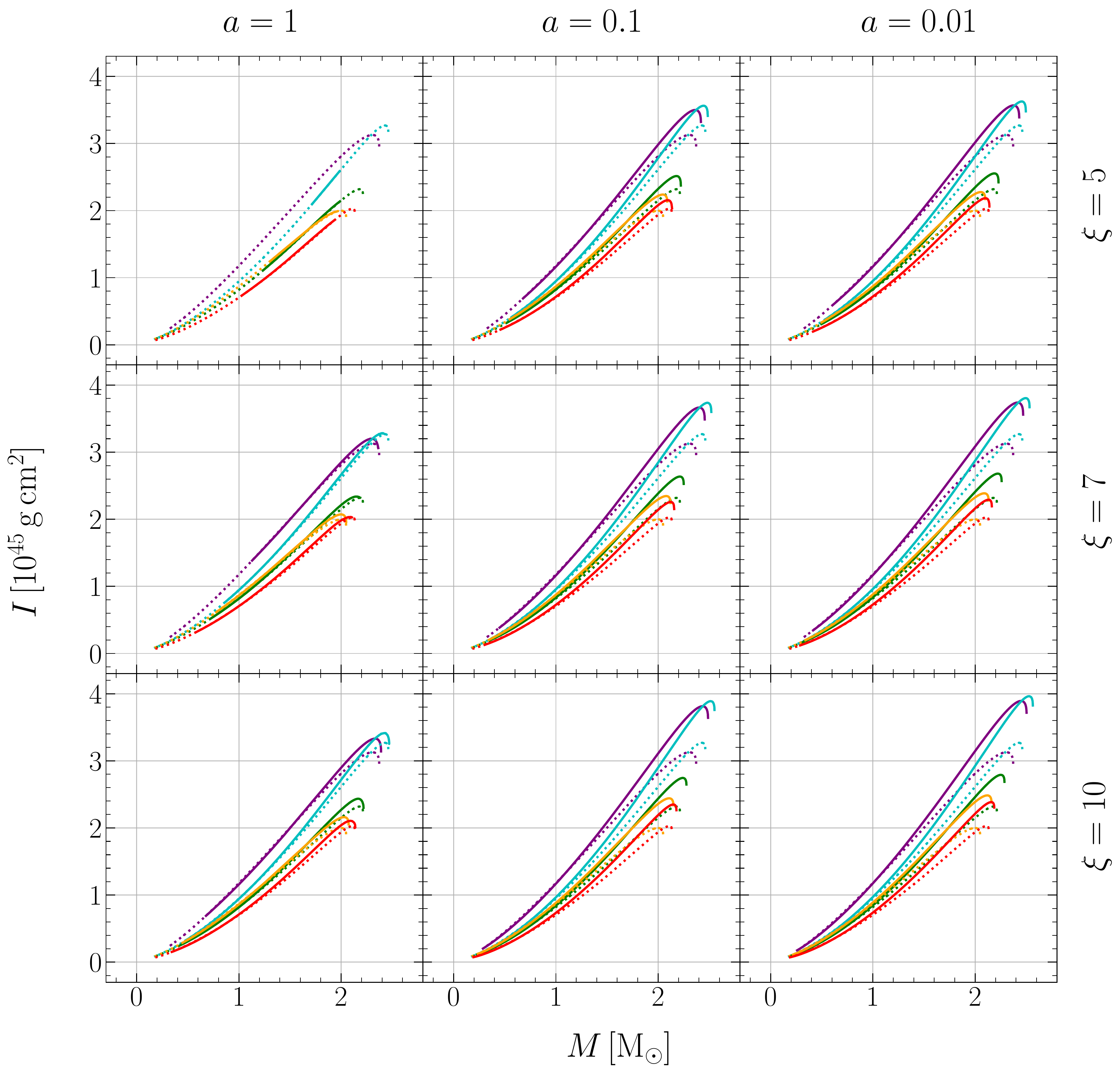}
 \caption{The moment of inertia $I$ versus the ADM mass $M$ for the same
 values of $\xi$ and $a$, as well as the same set of EOSs used in
 Fig.~\ref{fig3}. The solid curves are results of spontaneous
 scalarization, while the dotted curves are results from GR. }
 \label{fig5}
 \end{figure*}
 
The above calculations can be repeated with realistic EOSs of NSs. Dealing
with such EOSs hereafter, we take the characteristic length $l_0 = 10 \,
{\rm km}$, so the scalar mass is
\bea \label{eq:Phi:mass}
m_\Ph = \frac{a h}{2\pi l_0} \approx a \times 1.97 \times 10^{-11}\, {\rm eV}.
\label{dimlscalarmass}
\eea
Figure \ref{fig2} plots the relation $\xi_{\rm min}(a)$ for several
realistic EOSs: PAL1, MPA1, AP4, SLy4, and WFF1~\cite{Lattimer:2000nx}.
Except for PAL1, all the EOSs are chosen to simultaneously satisfy the
observations of two-Solar-mass pulsars~\cite{Antoniadis:2013pzd,
Cromartie:2019kug} and the tidal deformability from the binary NS merger
GW170817~\cite{TheLIGOScientific:2017qsa, Abbott:2018wiz, Abbott:2018exr}.
The EOS PAL1, representing a stiff EOS excluded by GW170817, is in the
list for comparison. For $a \rightarrow 0$, $\xi_{\rm min}$ is around $2$,
confirming the critical value of $\be \sim -4$ in the DEF theory via Eq.
\rf{destt} in the limit of a small scalar. We point out that $\xi_{\rm
min}$ increases significantly with $a$ when $a$ is large.

We also notice that when $a$ is from 0.01 to 1 in Eq.~\rf{dimlscalarmass}, the order of magnitude of the scalar mass $m_{\Ph}$
coincides with the values from the spin
measurement of superradiant black holes in X-ray
binaries~\cite{Brito:2015oca}. Superradiance for black holes with a light
scalar field is an interesting topic, which we will not go into detail in
this paper however.

\subsection{Numerical results of the nonlinear problem}
\label{sec:num}

Knowing the parameter space of $\xi$ and $a$ for spontaneous scalarization
from the study of the linearized scalar field, it is relatively
straightforward to numerically solve the group of nonlinear equations in
\rf{odegroup}. One thing that calls for attention is the numerical
singularity at $\rh = 0$. To avoid it, our numerical integrations start at
a small radius $\rh = \rh_{\rm min}$~\cite{Damour:1996ke}; it is also true
when numerically solving the linearized scalar equation from the center.
Series expansions of relevant functions at the center show that the values
of the functions at $\rh = \rh_{\rm min}$ are the same as their values at
$\rh = 0$ at least up to $O(\rh_{\rm min})$, but the derivatives of them
take corrections at $O(\rh_{\rm min})$. The two
derivatives used for starting numerical integrations are
\bea
\Ph'|_{\rh_{\rm min}} &=& \frac{\rh_{\rm min}}{3W}  {\left( 4\pi A^3 \frac{dA}{d\Ph}(\tilde \ep - 3 \tilde p) + A^3 \frac{dA}{d\Ph} U + \frac{1}{4} A^4 \frac{dU}{d\Ph} \right)}  \Bigg|_{\rh = \rh_{\rm min}} ,
\nonumber \\
\om'|_{\rh_{\rm min}} &=& \frac{16\pi \rh_{\rm min}}{5} A^4(\tilde \ep + \tilde p) \om \Big|_{\rh = \rh_{\rm min}} .
\eea
The fact that the solutions only depend on two independent input
parameters, the central pressure $\tilde p_c$ and the central scalar field
$\Ph_c$, is unaltered by slightly shifting the starting point of
integration.

Similar to solving the linearized scalar equation, the requirement of $\Ph
\rightarrow 0$ at infinity is achieved by using the shooting method. In
practice, among the two inputs, $\tilde p_c$ and $\Ph_c$, adjusting the
latter fulfills the requirement more efficiently. Varying $\tilde p_c$, on
the other hand, changes rapidly the parameters of the star, i.e., its
compactness and its mass.

Numerical results of spontaneous scalarization for various values of $\xi$
and $a$ in the regions above the curves in Fig.~\ref{fig2} have been
obtained. \Reply{Figure~\ref{figadd1} shows example solutions of the metric
components in the Jordan frame as functions of the radius $r$ defined in
Eq.~\rf{tranr} for NSs with $1.4 \, M_{\odot}$ and $2.0 \, M_{\odot}$.}

\Reply{Figures~\ref{fig3}--\ref{fig5} describe characteristic quantities of
NSs solved by varying the central pressure. In Fig.~\ref{fig3}, we choose
the values of the scalar field at the center and on the surface of the star
to display with respect to the ADM mass, signalling the extent of
spontaneous scalarization for NSs with different masses in the ST theories with three
representative values of $\xi$ and three representative values of $a$.} The
allowed interval of the ADM mass for spontaneous scalarization depends on
values of $\xi$ and $a$. The larger $\xi$ is and the smaller $a$ is, the
wider range $M$ covers for spontaneous scalarization. We point out that in
Fig.~\ref{fig3}, except for the upper left panel, unstable solutions come
up when approaching the upper onset of spontaneous
scalarization~\cite{Ramazanoglu:2016kul}. These unstable solutions are
removed from the plots.

To compare with GR solutions, the mass-radius relation and the change of
inertia moment with respect to $M$ are shown in Figs.~\ref{fig4} and
\ref{fig5} respectively. From the figures, we see that the deviations of
scalarized NSs from their counterparts in GR grow with $\xi$ and reduce
with $a$ in general. But for given values of $\xi$ and $a$, there is always
a scalarized NS having the same ADM mass and radius as its counterpart in
GR. When the mass of a NS is heavier than this special mass, its radius
increases once the star is scalarized, and vice versa. For this reason, a
heavy (light) scalarized NS acquires a larger (smaller) moment of inertia
compared to its counterpart in GR, as shown in Fig.~\ref{fig5}.
 
Our plots are qualitatively consistent with the results in
Refs.~\cite{Ramazanoglu:2016kul, Yazadjiev:2016pcb, Arapoglu:2019mun}. The
values of the scalar field at the center of the stars obtained in
Ref.~\cite{Ramazanoglu:2016kul} and the deviations from GR shown in the
$M$-$R$ and $I$-$M$ plots in Ref.~\cite{Yazadjiev:2016pcb} are relatively
larger than those in Figs.~\ref{fig3}, \ref{fig4} and \ref{fig5} here. The
difference is mainly from the specific forms of the nonminimal couplings used
in calculation. Compared to the exponential coupling function in Eq.
\rf{defa} adopted by Refs.~\cite{Ramazanoglu:2016kul, Yazadjiev:2016pcb},
the rational coupling function defined in Eq.~\rf{aeq} produces milder
spontaneous scalarization.

\section{Applications}
\label{sec:app}

In this section we study test-particle geodesics around scalarized NSs
and apply the lightlike geodesics to construct the pulse profiles of X-ray
pulsars with an illustrative model.

\subsection{Test particle geodesics}
\label{sec:app1}

Test particles around scalarized NSs follow trajectories different from
those in GR due to the distinctive metric solutions. Here we investigate
the difference by analytically considering the geodesics in a general
static spherical spacetime and numerically implementing several solutions
of scalarized NSs as an illustration. The geodesic equation holds in the
Jordan frame, so the motion of test particles will be studied using the
physical metric $\tilde g_{\mn}$.

Specifically, we define a new radial coordinate
\bea
r \equiv A(\varphi) \, \rh ,
\label{tranr}
\eea  
so that the line element in the Jordan frame can be obtained from Eq. \rf{metric} as
\bea
d\tilde s^2 = A^2 ds^2 &=& -g(r) \, dt^2 + f(r) \, dr^2 + r^2 d\th^2 
\nonumber \\
&& + r^2 \sin^2\th \Big( d\ph + (\om - \Om) \, dt \Big)^2 ,
\label{metric1}
\eea  
where $f$ and $g$ are related to $\mu$ and $\nu$ in
Eq.~\rf{metric} by
\beq
f = \left( \frac{d\rh}{dr} \right)^2 A^2 e^\mu , \quad
g = A^2 e^\nu .
\label{tranfg}
\eeq

\begin{figure}[h!]
 \includegraphics[width=0.8\linewidth]{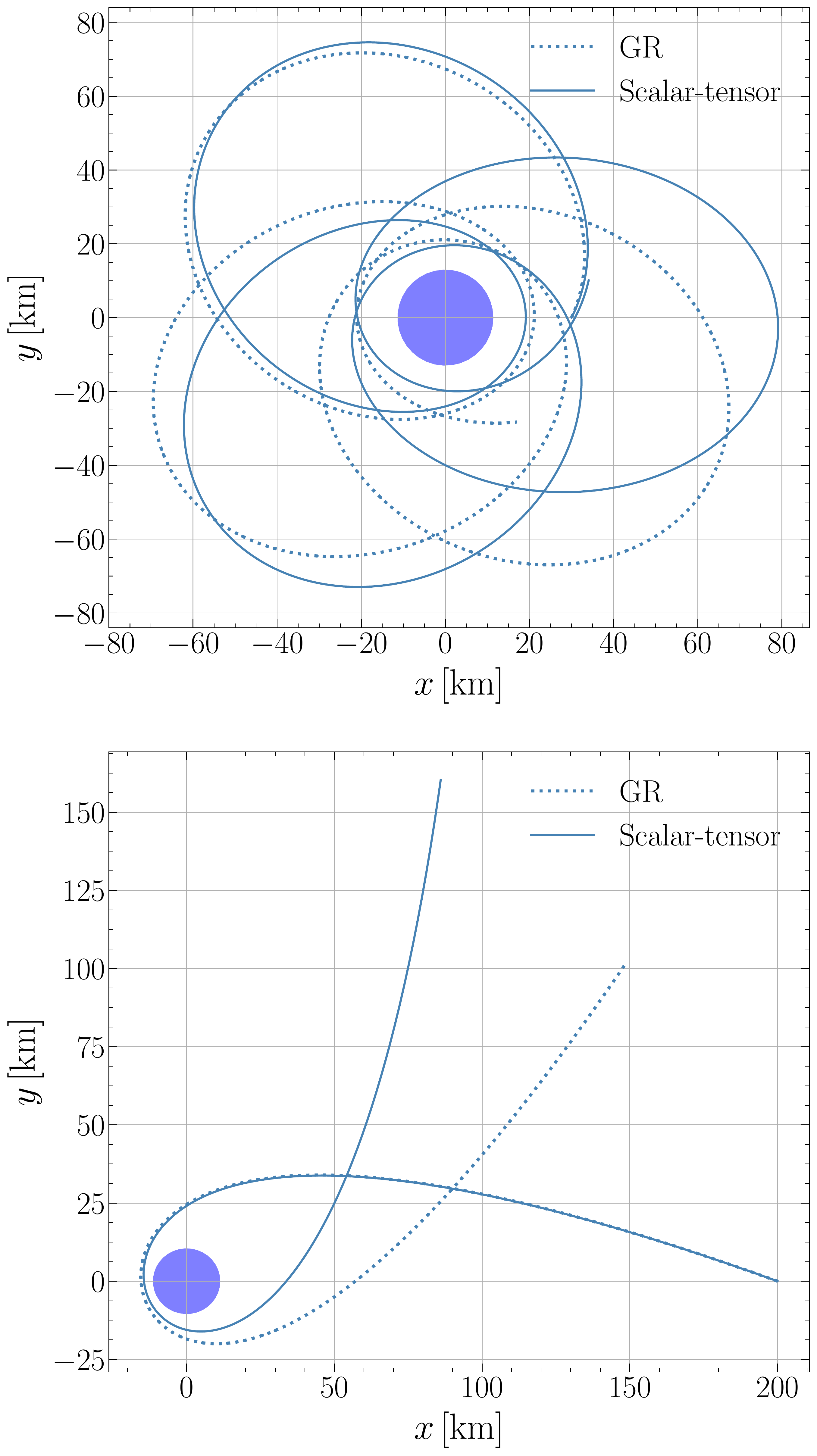}
  \caption{Timelike test-particle orbits around a NS of $M = 1.4 \,
  M_{\odot}$ with the EOS AP4. The GR orbits are given in dotted lines for
  comparison. The solid disk is the NS. The Cartesian coordinates are
  defined as $x = r\cos\ph, \, y = r\sin\ph$. ({\it Upper panel}) $\tilde E
  = 0.98,\, \tilde L = 4.5M$, and ({\it lower panel}) $\tilde E = 1, \,
  \tilde L = 4.5M$. }
\label{fig6}
\end{figure}

\def\arraystretch{1.6}
\begin{table*}
\caption{Properties of ISCOs and the maximal values of the angle $\psi$
defined in Fig. \ref{fig7} for several scalarized NSs solved with the EOS
AP4. Note that for $(\xi, a) = (5, 1)$, the ADM mass of scalarized NSs
cannot reach $2M_{\odot}$.  }
\begin{tabular}{l c c | c c c c c | c  }
  \hline   \hline
 \multicolumn{3}{p{4.8cm}|}{ \hspace{1.65cm} Parameters} & \multicolumn{5}{p{9.8cm}|}{ \hspace{4.2cm} ISCO properties} & \multicolumn{1}{p{2cm}}{ \hspace{0.2cm} Maximal $\psi$}   \\
   $(\xi, a)$ & $M\, (M_{\odot})$ & $R$ (km) & $r$ (km) & $ \hspace{1.2cm} \tilde E \hspace{1cm} $ &  $ \hspace{0.8cm} \tilde L \, (M) \hspace{1cm}$ & $\frac{d\ph}{dt}\, ({\rm ms}^{-1})$ & Grav. redshift & $\psi_{\rm max}$ \\
  \hline
   GR: $(0,0)$ & 1.40 & 11.42 & 12.4 & $\frac{2\sqrt{2}}{3} \approx 0.943$ & $2\sqrt{3} \approx 3.46$ & 9.87 & 0.816 & $2.15 \approx 123^\circ$ \\
   $(5, 0.01)$ & 1.40 & 11.46 & 12.2 & 0.943 & 3.60 & 10.3 & 0.805 & $2.13\approx 122^\circ$ \\
 
   $(5,1)$& 1.40 & 11.41 & 12.5 & 0.943 & 3.47 & 9.78 & 0.818 & $2.16\approx 124^\circ$ \\

   $(10,  0.01)$& 1.40 & 11.54 & 11.7 & 0.939 & 3.65 & 11.0 & 0.788 & $2.11 \approx 121^\circ$ \\

   $(10, 1)$& 1.40 & 11.39 & 13.3 & 0.948 & 3.59 & 9.05 & 0.828 & $2.15 \approx 123^\circ$ \\

   GR: $(0,0)$ & 2.00 & 11.00 & 17.7 & $\frac{2\sqrt{2}}{3} \approx 0.943$ & $2\sqrt{3} \approx 3.46$ & 6.91 & 0.816 & $2.89 \approx 165^\circ$ \\

   $(5, 0.01)$& 2.00 & 11.42 & 17.7 & 0.943 & 3.55 & 7.01 & 0.811 & $2.69 \approx 154^\circ$ \\
  
   $(10, 0.01)$& 2.00 & 11.63 & 17.2 & 0.941 & 3.60 & 7.36 & 0.800 & $2.61 \approx 150^\circ$ \\

   $(10,1)$& 2.00 & 11.26 & 18.3 & 0.944 & 3.49 & 6.59 & 0.823 & $2.78 \approx 159^\circ$ \\
  \hline
\end{tabular}
\label{tab1}
\end{table*}

To simplify the calculation, we will drop the $(\om-\Om)$ term in
Eq.~\rf{metric1} so that the metric $\tilde g_{\mn}$ is spherically
symmetric. Then, similar to the Schwarzschild spacetime, the geodesic
equation is completely integrable. The spherical symmetry also allows us to
take $\th = {\pi}/{2}$, and the first integrals are
\bea
g \frac{dt}{d\ta} &=& \tilde E , 
\nonumber \\
-g \left( \frac{dt}{d\ta} \right)^2 + f \left( \frac{dr}{d\ta} \right)^2 + r^2 \left( \frac{d\ph}{d\ta} \right)^2 &=& \ka ,
\nonumber \\
r^2 \frac{d\ph}{d\ta} &=& \tilde L ,
\label{trajeq}
\eea  
where the constants $\tilde E$ and $\tilde L$ are respectively the specific
energy and the specific angular momentum of the test particle, and the
constant $\ka$ takes $-1$ or $0$ for timelike or lightlike geodesics. The
motions can be conveniently studied by eliminating ${dt}/{d\ta}$ and
${d\ph}/{d\ta}$ to define an effective radial potential
\bea
 V_{\rm eff}(r) \equiv - \frac{1}{2} \left( \frac{dr}{d\ta} \right)^2 = \frac{1}{2f} \left( -\frac{ \tilde E^2 }{g} + \frac{\tilde L^2}{ r^2 } - \ka \right).
\label{veff}
\eea 
The radial coordinate $r$ of any orbit is confined to the range where
$V_{\rm eff}(r) \le 0$. Especially, circular orbits exist when $V_{\rm eff}
= 0$ and ${dV_{\rm eff}}/{dr} = 0$ \cite{Misner:1974qy}. In addition, the
circular orbit is stable (unstable) when ${d^2V_{\rm eff}}/{dr^2} > 0$
(${d^2V_{\rm eff}}/{dr^2} < 0$), while the inflection point ${d^2V_{\rm
eff}}/{dr^2} = 0$ determines the innermost stable circular orbit (ISCO).

Through Eq. \rf{tranfg}, $f$ and $g$ can be obtained from numerical
solutions of $\mu$ and $\nu$ for scalarized NSs, and then Eq. \rf{trajeq}
can be solved numerically for given $\tilde E$ and $\tilde L$. As examples,
Fig. \ref{fig6} shows a bound orbit and a scattering orbit around a
scalarized NS, while Table \ref{tab1} presents the properties of ISCOs for
several scalarized NSs. Table \ref{tab1} also contains the maximal values
of a quantity $\psi$, defined as the change of the angular coordinate $\ph$
for null geodesics from the surface of the star to infinity where the trajectories become straight lines. The angle $\psi$ is illustrate in Fig.
\ref{fig7} in the context of the X-ray radiation from a pair of hot spots
on a rotating NS (see the next subsection).

\subsection{Effects on pulse profiles from X-ray pulsars}
\label{sec:app2}

\begin{figure}[h!]
 \includegraphics[width=\linewidth]{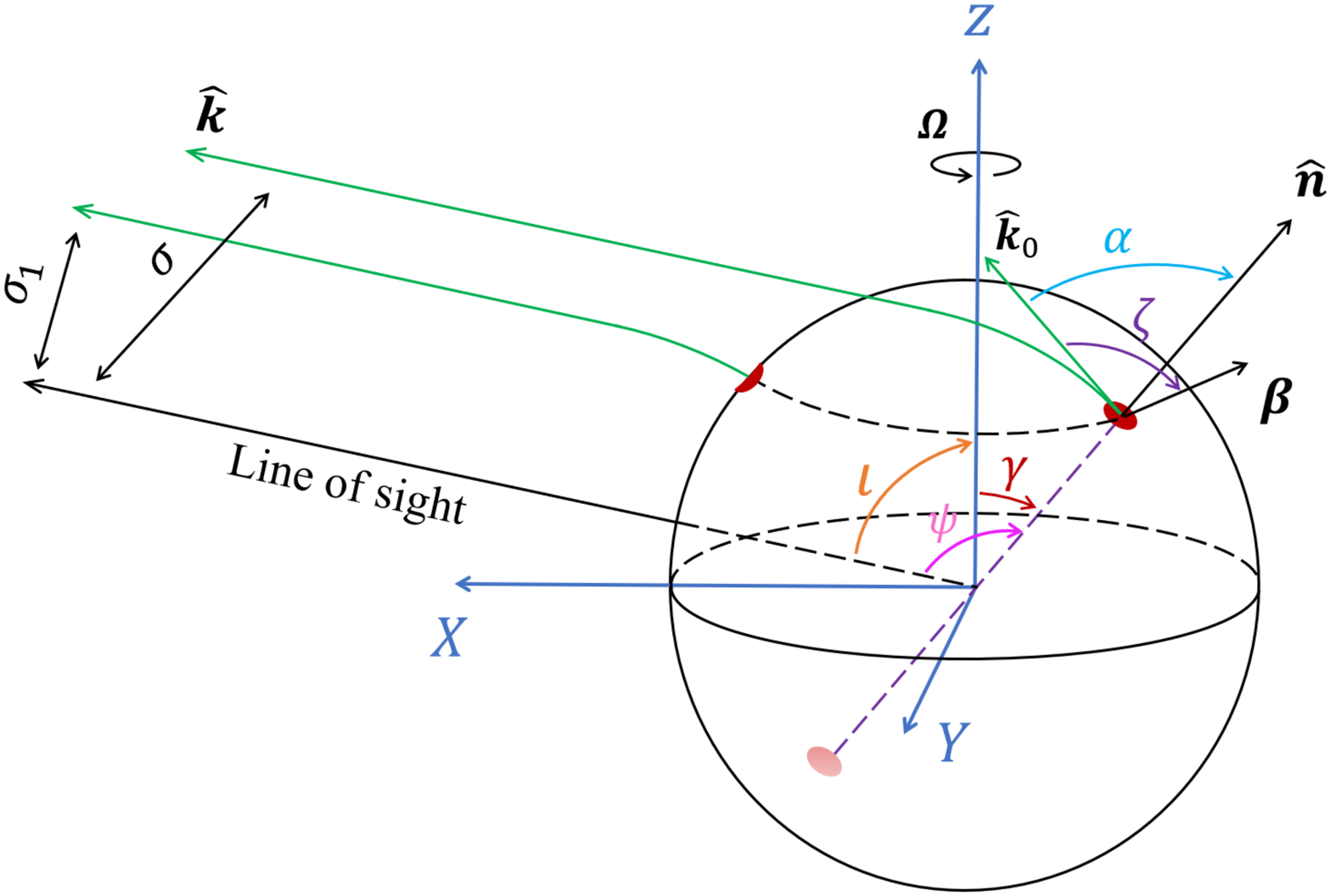}
  \caption{Schematic illustration for the X-rays emitted from a hot spot on
  a rotating NS and reaching the observer at infinity. The $Z$-axis is
  along the rotation axis of the NS, while the $X$-axis is set in the plane
  formed by the $Z$-axis and the line of sight. The upper green curve
  represents a general trajectory with its initial direction along a unit
  vector $\hat{\boldsymbol{k}}_0$ and its asymptotic direction along the
  line of sight, whose unit vector is $\hat{\boldsymbol{k}}$. The lower
  green curve specifically stands for the trajectory of the ray when the
  hot spot is in the $XZ$-plane and closest to the observer. The unit
  vector $\hat{\boldsymbol{n}}$ is pointing along the local radial
  direction, and the vector $\boldsymbol{\be}$ is the velocity of the hot
  spot in the local static frame. Five relevant angles are indicated: the
  angle $\io$ between the line of sight and the $Z$-axis, the colatitude
  $\ga$ of the hot spot in the $XYZ$-frame, the angle $\psi$ between the
  line of sight and $\hat{\boldsymbol{n}}$, the angle $\al$ between
  $\hat{\boldsymbol{k}}_0$ and $\hat{\boldsymbol{n}}$, and the angle $\ze$
  between $\hat{\boldsymbol{k}}_0$ and $\boldsymbol{\be}$.}
\label{fig7}
\end{figure}

When radiation is emitted from a scalarized NS, the observed
\Reply{bolometric} flux takes modification compared to the case of GR due
to the difference in the bending of light by the NS spacetime. Following
the method of \citet{Silva:2018yxz}, we numerically calculate the observed
\Reply{bolometric} fluxes for the X-ray radiation emitted by a pair of hot
spots on the surfaces of the NSs. Figure \ref{fig7} illustrates the
physical picture in consideration, and various notations are explained in
the caption.

We follow the assumptions in
Ref.~\cite{Silva:2018yxz} in calculating the observed flux. To ease the reading, we briefly review them
here. First, the radiative hot spots are assumed to have infinitesimal
areas sitting oppositely on the surface of the NS at two poles. The
specific intensities of the radiation at two spots are assumed to have the
same dependence on the energy and the emitting direction in their locally
comoving frames. For simplicity in the demonstration, the radiation is
additionally presumed to be isotropic, leaving the specific intensity a
function of the energy alone. The model of isotropic radiation works if the
hot spots are blackbodies and sit in vacuum. However, for realistic hot
spots on the surface of a NS, the presence of a strong magnetic field and
the Compton scattering in the magnetosphere of the NS generate nontrivial
angular patterns of the radiation even if the hot spots themselves can be
approximated as blackbodies (e.g., see Refs.~\cite{Meszaros:1992tf,
1991ApJ...376..161M, Ozel:2001jq, Poutanen:2008pg}). Therefore, the
isotropic assumption must be replaced by a more realistic angular
distribution of the radiation when the theoretical predictions are to be
confronted against data from X-ray pulsar observations
\cite{Poutanen:2006hw}. Second, the trajectories of photons are assumed to
follow Eq.~\rf{trajeq} with $\ka = 0$. In doing so, the effect from the
rotation of the NS on the spacetime is neglected, namely that we have
assumed a spherically symmetric curved spacetime. Also, if the X-ray pulsar
lives in a binary system, we are neglecting the effect on spacetime caused
by its companion star. Finally, the distance from the observer to the NS is
assumed to be large enough so that it can be mathematically treated as
infinity. These assumptions can be relaxed when
necessary.

With the above assumptions and the relevant angles defined in Fig.
\ref{fig7}, we summarize the derivation for the observed flux at infinity,
as demonstrated in Ref.~\cite{Silva:2018yxz}, into the following steps.
\begin{enumerate}
    \item Using the specific intensity at infinity, $I(E,
    \hat{\boldsymbol{k}})$, which is a function of the observed energy $E$,
    and the unit vector $\hat{\boldsymbol{k}}$ along the observed direction
    of the ray, the observed differential flux can be expressed as
    \bea
    dF = I(E, \hat{\boldsymbol{k}}) \, dE \, d\Om = I(E, \hat{\boldsymbol{k}} ) \, dE \, \frac{\si d\si d\la}{D^2} , 
    \label{flux}
    \eea 
    where $d\Om$ is the differential solid angle formed by light rays
    coming to the observer with the impact parameter ranging from $\si$ to
    $\si + d\si$ and the azimuth angle around $\hat{\boldsymbol{k}}$
    ranging from $\la$ to $\la + d\la$. The distance $D$ from the observer
    to the NS is assumed to be large.

    \item To express Eq. \rf{flux} in terms of quantities at the emission
    point, the fact that the quantity ${I}/{E^3}$ is conserved along light
    rays is useful. Denoting the specific intensity in the local static
    frame at the emission point as $I_0 (E_0, \hat{\boldsymbol{k}}_0)$,
    which is a function of the emitting energy $E_0$, and the unit vector
    $\hat{\boldsymbol{k}}_0$ along the emitting direction, gravitational
    redshift leads to
    \bea
    I = \left( \frac{E}{E_0} \right)^3 I_0 =  g^{\frac{3}{2}}(R) \, I_0 .
    \eea
    The impact parameter $\si$ is related to the angle $\psi$, \Reply{defined as the change of the angular coordinate $\ph$ for null geodesics in Eq.~\rf{trajeq} from the surface of the star to the distant observer (see Fig.~\ref{fig7})}, by
    integrating the trajectory equation, namely
    \bea
    \psi(\si) = \Reply{\int_R^\infty \left|\frac{d\ph}{dr} \right| dr =} \si \int_R^\infty \frac{\sqrt{fg} }{r^2} \frac{1}{ \sqrt{ 1 - \frac{\si^2 g}{r^2} } } dr, 
    \label{psisi}
    \eea
    where the specific energy $\tilde E$ and the specific angular momentum
    $\tilde L$ of the photon \Reply{in Eq.~\rf{trajeq}} have been eliminated by $\si$ using the
    relation
    \bea
    \si = \frac{\tilde L}{\tilde E} .
    \eea
    Therefore, the observed differential flux in terms of the quantities at
    the emission point is
    \bea
    dF &=&  g^{2}(R) \, I_0 \, dE_0 \, \frac{\si(\psi) d\psi d\la}{D^2} \, \frac{d\si}{d\psi} 
    \nonumber \\
    &=& \frac{ g^2(R) \, I_0 \, dE_0 \, dS }{D^2 \, R^2} \, \frac{\si(\psi)}{ \sin\psi} \, \frac{d\si}{d\psi} ,
    \label{flux1}
    \eea
    where the function $\si(\psi)$ is the inverse of the function
    $\psi(\si)$ in Eq. \rf{psisi}, and $dS = R^2 \sin\psi d\psi d\la$ is
    the differential surface area in the local static frame at the emission
    spot.

    \item Only considering the kinematic effect caused by the rotation of
    the NS, a local Lorentz boost is requisite for bringing $I_0, \, E_0$
    and $dS$ into the locally comoving frame at the spot. Their
    transformations under a boost are (e.g., see
    Ref.~\cite{Rybicki:2004hfl})
    \beq
    dS = \de \, dS', \quad
    E_0 = \de \, E_0' ,  \quad
    I_0 = \de^3 I_0',
    \label{lt}
    \eeq
    where $\de$ is the relativistic Doppler factor, which depends on the
    boost speed $\be$ and \Reply{the angle $\ze$ between the boost velocity $\boldsymbol{\be}$ and the direction of the radiation $\hat{\boldsymbol{k}}_0$} via \cite{Rybicki:2004hfl}
    \bea
    \de = \frac{ \sqrt{1-\be^2} }{1-\be \cos\ze} .
    \eea
    In our case, $\be$ and $\cos\ze$ can be calculated by \cite{Poutanen:2006hw}
    \bea
    \be &=& \frac{R\Om \sin\ga}{ \sqrt{-g(R)} } ,
    \nonumber \\
    \cos\ze &=& -\frac{\sin\al \sin\io \sin \big(\Om t_0\big)}{\sin\psi} ,
    \eea
    where $t_0$ is the coordinate time when the ray is emitted and has been
    set to zero when the spot is closest to the observer. \Reply{The angle
    $\ga$ is the colatitude of the hot spot; the angle $\al$ spans from the
    local radial direction of the spot to the direction of the radiation;
    and $\io$ is the inclination angle of the distant observer (see
    Fig.~\ref{fig7}).} Substituting Eq.~\rf{lt} into Eq.~\rf{flux1}, we get
    \bea
    dF &=& \frac{ \de^5 \, g^{2}(R) \, I_0' \, dE_0' \, dS' }{D^2 \, R^2} \, \frac{\si(\psi)}{ \sin\psi} \, \frac{d\si}{d\psi} ,
    \nonumber \\
    &=& \frac{ \de^5 \, g(R) \, I_0' \, dE_0' \, dS' }{D^2} \, \frac{\sin\al \cos\al}{ \sin\psi} \, \frac{d\al}{d\psi} ,
    \label{flux2}
    \eea
    where the relation
    \bea
    \sin\al = \frac{\si \sqrt{g(R)} }{R} ,
    \eea
    has been used to eliminate $\si$ using $\al$.

    \item Finally, as the star rotates, the angle $\psi$
    changes according to
    \bea
    \cos\psi = \cos\io \cos\ga + \sin\io \sin\ga \cos \big(\Om t_0\big) ,
    \label{psit}
    \eea
    which, interpreted as a function of $t_0$ and combined with Eq.
    \rf{flux2}, gives the relation between the observed flux $dF$ and the
    emission time $t_0$ for given values of $\io$ and $\ga$. However, the
    observer time $t$, is delayed compared to $t_0$ by the traveling time
    of the photon
    \bea
    t(\si) - t_0 = \int_R^\infty \sqrt{ \frac{ f} { g} } \frac{1}{ \sqrt{1 - \frac{\si^2 g}{r^2} } } dr ,
    \eea   
    which diverges as the trajectory extends to infinity mathematically.
    For our purpose, a relative time delay can be defined by
    \bea
    \de t \,(\si) &\equiv& t(\si) - t_0 - \int_R^\infty \sqrt{ \frac{ f} { g} } dr 
    \nonumber \\
    &=&  \int_R^\infty \sqrt{ \frac{ f} { g} }  \left( \frac{1}{ \sqrt{1 - \frac{\si^2 g}{r^2} } } - 1 \right) dr ,
    \label{detsi}
    \eea
    which turns out to be finite. Assuming the impact parameter is $\si_1$
    when the spot is closest to the observer as shown in Fig. \ref{fig7},
    then by resetting the observer time as $t' \equiv t(\si) - t(\si_1)$,
    the emission time $t_0$ can be expressed by the new observer time $t'$
    as
    \bea
    t_0 = t' - \de t\, (\si) + \de t\, (\si_1) .
    \label{timetr}
    \eea  
    Now by replacing the emission time $t_0$ with the observer time $t'$
    using Eq. \rf{timetr}, the observed flux $dF$ in Eq. \rf{flux2} is a
    function of the observer time $t'$.

\end{enumerate}

Following \citet{Silva:2018yxz}, we integrate Eq. \rf{flux2} over $dE_0'$
under the isotropic assumption, and numerically calculate the normalized
flux
\bea
F \equiv \frac{D^2 \int dF}{ dS' \int I'_0 \, dE'_0  } = \de^5 g(R)  \frac{\sin\al \cos\al}{ \sin\psi} \frac{d\al}{d\psi} .
\label{flux4}
\eea
Equations \rf{flux2} and \rf{flux4} are essentially the generalized version
of the results in Ref.~\cite{Silva:2018yxz} for any static spherical
spacetime.

\begin{figure}[h!]
 \includegraphics[width=\linewidth]{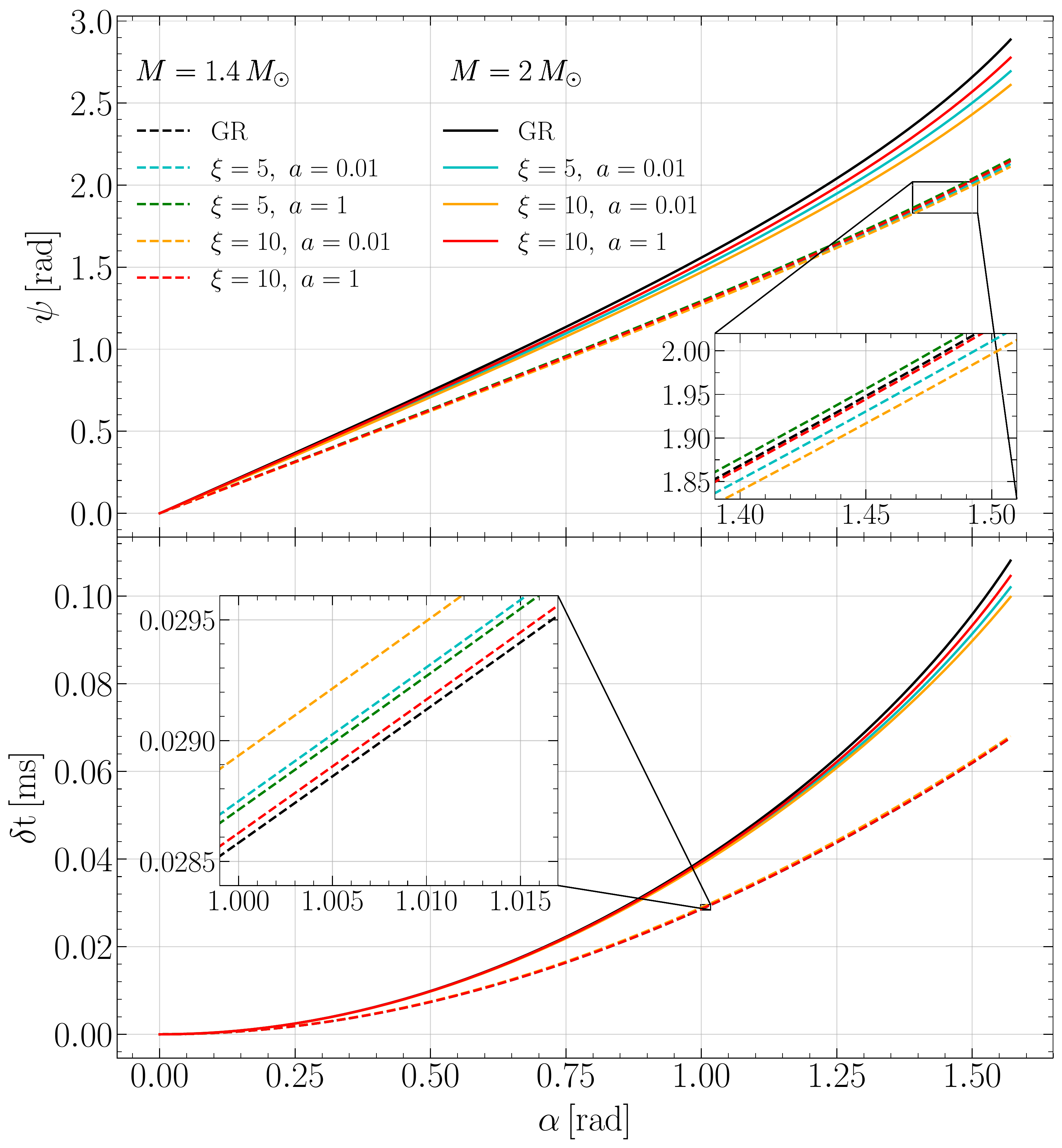}
  \caption{({\it Upper panel}) $\psi $ versus $\al$, and ({\it lower
  panel}) $\de t$ versus $\al$. The solid curves are for the $2\,M_{\odot}$
  NSs in Table~\ref{tab1}, and the dashed curves are for the $1.4 \,
  M_{\odot}$ NSs in Table~\ref{tab1}. NSs with the same mass but different
  values of $\xi$ and $a$ are distinguished by colors. The results in GR
  are plotted in black for comparison. }
\label{fig8}
\end{figure}

\begin{figure*}
 \includegraphics[width=0.9\linewidth]{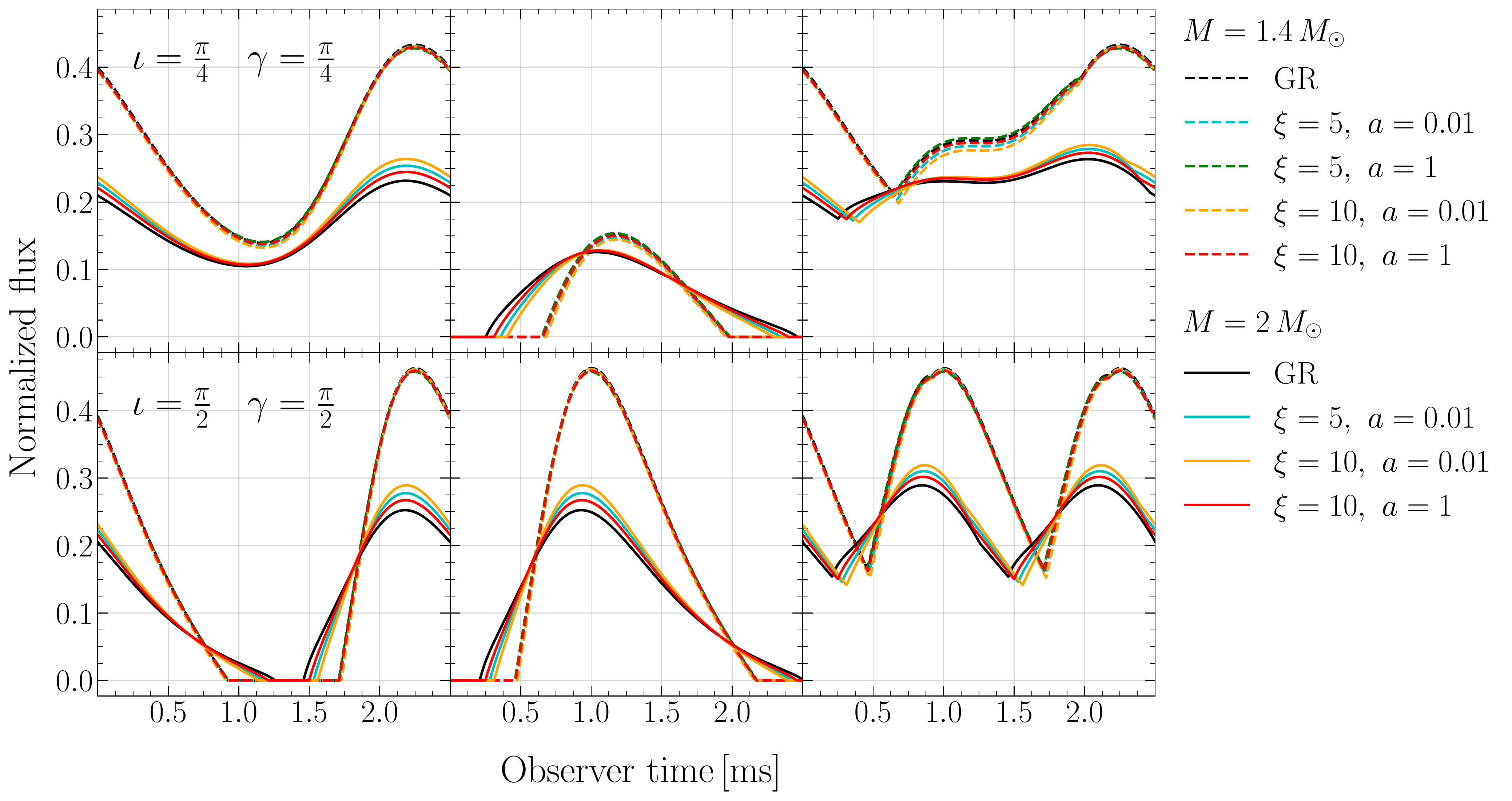}
  \caption{The X-ray pulse profiles for NSs listed in Table \ref{tab1} with
  $\io = {\pi}/{4}, \, \ga = {\pi}/{4}$ (upper panels) and $\io =
  {\pi}/{2}, \, \ga = {\pi}/{2}$ (lower panels). The angular frequency of
  the NSs is taken to be $ 400 \, {\rm Hz}$. ({\it Left panels}) The pulse
  profiles of the spots at colatitude $\ga$. ({\it Middle panels}) The
  pulse profiles of the antipodal spots at colatitude $\pi - \ga$. ({\it
  Right panels}) The pulse profiles of the pair of spots summed. The solid
  curves are for $2 \, M_{\odot}$ NSs, and the dashed curves are for $1.4
  \, M_{\odot}$ NSs. NSs with the same mass but different values of $\xi$
  and $a$ are distinguished by colors. }
\label{fig9}
\end{figure*}

In Fig.~\ref{fig8}, $\psi$ calculated with Eq. \rf{psisi} and $\de t$
calculated with Eq. \rf{detsi} are plotted as functions of $\al$, for the
scalarized NSs listed in Table \ref{tab1}. In Fig.~\ref{fig9}, by choosing
two sets of $(\io, \, \ga)$ for the hot spots and assuming the rotation
frequency of the NS to be $400\, {\rm Hz}$, their fluxes calculated with
Eq. \rf{flux4} are displayed with respect to the observer time $t'$.

From Fig. \ref{fig8}, we can see that the deviations from GR become
perceivable for scalarized NSs with $2 \, M_{\odot}$ as $\al$ increases
from 0 to ${\pi}/{2}$. But for NSs with $1.4 \, M_{\odot}$, the deviations
from GR can only be seen when the graphs are zoomed in. This is the reason
why the X-ray pulse profiles for the NSs with $1.4 \, M_{\odot}$ in Fig.
\ref{fig9} are mostly indistinguishable. We do notice that by summing the
fluxes from the spot and from the antipodal spot, the degeneracy weakens a
bit (the dashed profiles in the upper right panel in Fig. \ref{fig9}).
Another feature in the X-ray pulse profiles is that the flux from a single
spot might be cut off if the spot rotates to a position where $\psi$
calculated from Eq. \rf{psit} is greater than the maximal values presented
in the last column of Table~\ref{tab1}. However, because $\psi_{\rm max}$
is generally greater than ${\pi}/{2}$, when taking the pair of spots into
consideration, there is no cut-off region in the profiles (right panels in
Fig. \ref{fig9}).

\section{Summary}
\label{sec:disc}

ST theories can possess distinct solutions for NSs from those of GR through
spontaneous scalarization. Taking the nonminimal coupling between the
scalar field and gravity from inflationary models, we study a class of
specific massive ST theories described by the action \rf{eq:action} in this
work. The field equations in both the Jordan frame and the Einstein frame
are shown, and with the configuration of a slowly rotating NS the field
equations are simplified to the set of ODEs in Eq.~\rf{odegroup}. A match
of the theory \rf{eq:action} with the DEF theory with a mass term in the
regime of linearized scalar is established via Eq.~\rf{destt}. Then the
linearized scalar equation is investigated to obtain the theory parameter
space for spontaneous scalarization. The results, plotted in Fig.
\ref{fig2}, are also indicative for the DEF theory with a mass term through
the match in Eq.~\rf{destt}.

Numerical solutions for the full set of the ODEs in Eq.~\rf{odegroup} have
been obtained, and geodesics around the scalarized NSs have been calculated
as an application. As preparation for putting the theory into tests, we
calculated the mass-radius relation (Fig. \ref{fig4}), the
moment-of-inertia--mass relation (Fig. \ref{fig5}), and the X-ray pulse
profiles (Fig. \ref{fig9}) to compare with those predicted in GR. An
interesting observation is that there is a special mass at which NSs are
almost identical in the theory \rf{eq:action} and in GR. Though depending
on the theory parameters $\xi$ and $a$, as well as the EOS, the special
mass for a wide range of the theory parameters and most of the EOSs that we
use is around $1.5 \, M_{\odot}$. This makes distinguishing the theory
\rf{eq:action} and GR difficult using observations of NSs with masses
around $1.5 \, M_{\odot}$. In fact, due to the substantial uncertainties in
measuring NS radii and moments of inertia, the theory \rf{eq:action} with
$\xi$ as large as 10 and $a$ as small as 0.01 still produces $M$-$R$ and
$I$-$M$ relations consistent with the observations as long as the EOSs used
are not excluded by observations interpreted under GR.

The difficulty of distinguishing the theory \rf{eq:action} and GR using
observations might be solved with the ongoing Neutron star Interior
Composition Explorer (NICER) mission \cite{Ray:2017tzb} as well as the
revolutionary gravitational wave (GW) detecting technology. By employing
sophisticated techniques during data analysis, small differences in X-ray
pulse profiles might be distinguishable. Following
Ref.~\cite{Silva:2018yxz}, we have demonstrated constructing the X-ray
pulse profile of a slowly rotating NS which possesses a general static
spherical metric. To produce realistic X-ray pulse profiles for NICER to
test, an integration of Eq.~\rf{flux2} that takes the angular pattern of the
radiation into consideration is essential. The effect of NS rotation on the
spacetime might also make a small difference on the pulse profile. Those
are aspects worthy of future study. As for tests with GW observations from
coalescences of binary NS systems, our solution of a single NS is only the
prelude. A proper adaption of the post-Newtonian approximation developed
for the massive Brans-Dicke theory in Ref.~\cite{Alsing:2011er} is desired.
This lies outside the scope of the current work and we leave it as a
direction for future study.

\acknowledgments

\Reply{It is a pleasure to thank the anonymous referee for helpful
comments. We are grateful to Zhoujian Cao, Kohei Inayoshi, and Jiayin Shen
for discussions.}
This work was supported by the National Natural Science Foundation of China
(11975027, 11991053, 11721303), the Young Elite Scientists Sponsorship
Program by the China Association for Science and Technology (2018QNRC001),
the Max Planck Partner Group Program funded by the Max Planck Society, and
the High-performance Computing Platform of Peking University.
It was partially supported by the Strategic Priority Research Program of
the Chinese Academy of Sciences through the Grant No. XDB23010200.
R.X. is supported by the Boya Postdoctoral Fellowship at Peking University.

\appendix
\section{Series expansion at $x=0$ for the linearized scalar equation}
\label{app1}

Let us investigate the behavior of $\Ph$ near the center of the star if it
satisfies Eq. \rf{linphdim} with $u = {3}/{2}$ and $y, \, v$ given by
Eq.~\rf{mpsol}. Substituting the series expansion
\bea
\Ph = \sum\limits_{n=0}^{n=\infty} \frac{\Ph_n}{n!} x^n ,
\eea  
into Eq.~\rf{linphdim}, and keeping in mind $\Ph_1 = 0$ due to the singular factor ${2}/{x}$ in front of $\Ph_x$, we find  
{ \allowdisplaybreaks
\bea
 x^0: \quad &&\Ph_2 + 2 \Ph_2 + \left( 3\xi - a^2 -  \frac{6\xi(3 \et - 1)}{2 (1 - \et)} \right) \Ph_0 = 0, 
\nonumber \\
 x^1: \quad &&\Ph_3 + \Ph_3  = 0,
\nonumber \\
 x^2: \quad &&\frac{\Ph_4}{2!} - \Ph_2 + \frac{\Ph_4}{3} - \frac{1}{2!} C_2^1 \, z_1 \, \Ph_{2}   
\nonumber \\
&& + (3\xi - a^2) \frac{\Ph_2}{2!} - 3\xi \left( \frac{3 \et - 1}{2 (1 - \et)} \Ph_2 -\frac{\et}{2 (1-\et)^2}\Ph_0 \right) = 0,
\nonumber \\
 x^3: \quad &&\frac{\Ph_5}{3!} - \Ph_3 + \frac{\Ph_5}{12} - \frac{1}{3!} C_3^1 \, z_1 \, \Ph_{3}  
\nonumber \\
&& + (3\xi - a^2) \frac{\Ph_3}{3!} - \frac{\xi(3 \et - 1)}{2 (1 - \et)} \Ph_3 = 0,
\eea
and generically, we have
\bea
 x^n: \quad &&\frac{\Ph_{n+2}}{n!} - \frac{\Ph_{n}}{(n-2)!} + \frac{2\Ph_{n+2}}{(n+1)!} - \frac{1}{n!} \sum\limits_{k=1}^{n} C_n^k \, z_k \, \Ph_{n-k+1}  
\nonumber \\
&& + (3\xi - a^2) \frac{\Ph_n}{n!} - \frac{6\xi}{n!} \sum\limits_{k=1}^{n} C_n^k \, v_k \, \Ph_{n-k} = 0 ,
\eea
}
where $C_n^k$ denotes the binomial coefficient, and $z_k$ and $v_k$ are the
$k$-th derivatives of $z \equiv \left( \frac{5}{2} - v \right) x $ and $v$
at $x = 0$. From the second equation in \rf{mpsol} we see that $v$ is an
even function of $x$, which indicates $z$ to be odd. Therefore, the even
and odd coefficients decouple in the recurrence relation of $\Ph_n$. In
addition, all the odd coefficients vanish due to $\Ph_3 = 0$, and all the
even coefficients are proportional to $\Ph_0$ through the linear recurrence
relation. In conclusion, the solution of $\Ph$ near the center of the star is even and fixed up to
an overall scaling constant which can be conveniently chosen as $\Ph_0$. 

The above conclusion is drawn for the toy EOS, but it is verified to be true for realistic EOSs by observing that there is no physical solution of odd $\Ph$ directly using numerical calculation.

\bibliography{refs}

\end{document}